\documentclass[preprint,preprintnumbers,a4paper,aps,superscriptaddress]{revtex4}
\usepackage[dvips]{graphicx}
\usepackage{amsmath}

\begin{document}

\title{Energies of $B_s$ meson excited states --- a lattice study}

\collaboration{UKQCD Collaboration}
\noaffiliation
\author{J. Koponen}
\email{jonna.koponen@helsinki.fi}
\affiliation{Department of Physics and
Helsinki Institute of Physics,
P.O. Box 64, FIN--00014 University of Helsinki,Finland}

\date{\today}
\preprint{HIP-2007-44/TH}

\begin{abstract}
This is a follow-up to our earlier work on the energies  and radial distributions
of heavy-light mesons. The heavy quark is taken to be static (infinitely heavy)
and the light quark has
a mass about that of the strange quark. We now concentrate on the energies of the
excited states with higher angular momentum and with a radial node.
A new improvement is the use of hypercubic blocking in the time direction.

The calculation is carried out with dynamical fermions on a $16^3 \times 32$
lattice with a lattice spacing $a\approx 0.1$~fm generated using a
non-perturbatively improved clover action.

In nature the closest equivalent of this heavy-light system is the $B_s$ meson,
which allows us to compare our lattice calculations to experimental results (where
available) or to give a prediction where the excited states, particularly P-wave states,
should lie. We pay special attention to the spin-orbit splitting, to see which
one of the states (for a given angular momentum L) has the lower energy.
An attempt is made to understand these results in terms of the Dirac
equation.

\end{abstract}

\maketitle
  
\section{Motivation}

There are several advantages in studying a heavy-light system on a lattice.
First of all, the lattice calculations are relatively easy to do, and it allows us
to do these with QCD from first principles. Our meson is much more simple
than in true QCD: one of the quarks is static (infinitely heavy) with the light
quark ``orbiting'' it. This makes it very beneficial for modelling. On the lattice
an abundance of data can be produced, and we know which state we are measuring.
In contrast, the physical states can be a mixture of two or more configurations,
but on the lattice this complication is, mostly, avoided. Even so, our results on the
heavy-light system can still be compared to the $B_s$ meson experimental
results.

\section{Measurements and lattice parameters}

We measure the energies of both angular and first radial excitations
of heavy-light mesons.
Since the heavy quark spin decouples from any description of the
configurations we may label the states
as $\mathrm{L}_{\pm}=\mathrm{L}\pm\frac{1}{2}$, where L is the
orbital angular momentum
and $\pm\frac{1}{2}$ refers to the spin of the light quark.

The measurements are done on $16^3\!\times\! 32$ lattices using two degenerate quark
flavours. The lattice configurations were generated by the UKQCD Collaboration using
lattice action parameters \mbox{$\beta = 5.2$}, $c_\textrm{SW} = 2.0171$ and three
different values for the hopping parameter $\kappa$ (see Table~\ref{LatParams}). The
three different lattices are referred to here as ``DF3'', ``DF4'' and ``DF5''.
Each of them has a slightly different lattice spacing ($a$) and a different light
quark mass ($m_q$). Our main results are measured on the ``DF3'' lattice, because
the light quark mass is very close to the strange quark mass.
More details of the lattice configurations used in this study can be found in
Refs.~\cite{Allton,PRD69}. Because our light quarks are heavier than true \textit{u}
and \textit{d} quarks, we have $m_{\pi}$ ranging from 730~MeV (``DF3'') to 400~MeV (``DF5'').
Two different levels of fuzzing (2 and 8 iterations of conventional fuzzing) are
used in the spatial directions to permit a cleaner extraction of the excited states.

\begin{table}[b!]
\begin{center}
\begin{tabular}{|c|c|c|c|c|c|c|}
\hline
\hline
   &$\kappa$&$r_0/a$&$a$ [fm] (approx.)&$m_q/m_s$ (approx.)&$r_0m_\pi$& No. of configs.\\
\hline
DF3&0.1350  &$4.754(40)^{+2}_{-90}$&0.110 &1.1      &1.93(3) & 160 \\
DF4&0.1355  &$5.041(40)^{+0}_{-10}$&0.104 &0.6      &1.48(3) & 119 \\
DF5&0.1358  &$5.32(5)$             &0.099 &0.3      &1.06(3) & 139 \\
\hline
\hline
\end{tabular}
\end{center}
\caption{Lattice parameters (from~\cite{Allton}). The Sommer scale parameter
can be taken to be $r_0=0.525(25)$~fm and $m_s$ is the \textit{s} quark mass.}
\label{LatParams}
\end{table}

\section{2-point correlation function}

\begin{figure}[b]
\centering
\includegraphics*[width=0.31\textwidth]{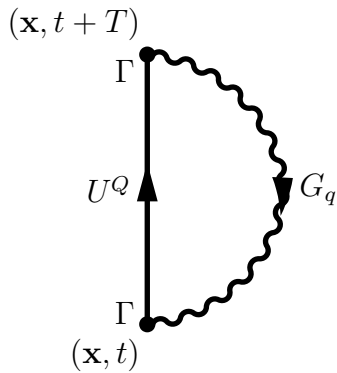}
\caption{
Two-point correlation function.
}
\label{Fig:Corr}
\end{figure}

The 2-point correlation function (see Fig.~\ref{Fig:Corr}) is defined as
\begin{equation}
\label{2point}
C_2(T)=\langle P_t\Gamma G_q(\mathbf{x},t+T,t)P_{t+T}
\Gamma^{\dag}U^Q(\mathbf{x},t,t+T)\rangle \  ,
\end{equation}
where $U^Q(\mathbf{x},t,t+T)$ is the heavy (infinite mass)-quark propagator
and $G_q(\mathbf{x},t+T,t)$ the light anti-quark propagator. $P_t$
is a linear combination of products of gauge links at time $t$
along paths $P$ and $\Gamma$ defines the spin structure of the operator.
The $\langle ...\rangle$ means the average over the whole lattice.
A detailed discussion of lattice operators for orbitally excited mesons
can be found in~\cite{Lacock}. In this study, the same operators are used
as in~\cite{MVR}.
The energies ($m_i$) and amplitudes ($a_i$) are extracted by fitting the $C_2$
with a sum of exponentials,
\begin{equation}
\label{C2fit}
[C_2(T)]_{f_1,f_2}\approx
\sum_{i=1}^{N_{\textrm{max}}}a_{i, f_1}\mathrm{e}^{-m_i T}a_{i, f_2},\;
\textrm{where $N_{\textrm{max}}=2\textrm{ -- }4$, $T\leq 14$}.
\end{equation}
The fit is a simple least squares fit.
In most of the cases 3 exponentials are used to try to ensure the first
radially excited  states are not polluted by higher states.
Also 2 and 4 exponential fits are used to cross-check the results wherever
possible. Indices $f_1$ and $f_2$ denote the amount of fuzzing used at the
vertices and both of them take two values, $f_1 = $F1, F2
and $f_2 = $F1, F2, where (F1=2 iterations and F2=2+6 iterations).
For S and P$_-$ states we have alternative operators (see~\cite{MVR}), so we
get a 5 by 5 matrix (5 paths, because one operator has two choices, F1 and F2,
and the other operator has three choices, local, F1 and F2) instead of just a
2 by 2 matrix (2 paths) given by the fuzzing choices.

\section{Smeared heavy quark}

\begin{figure}
\centering
\includegraphics[width=0.475\textwidth]{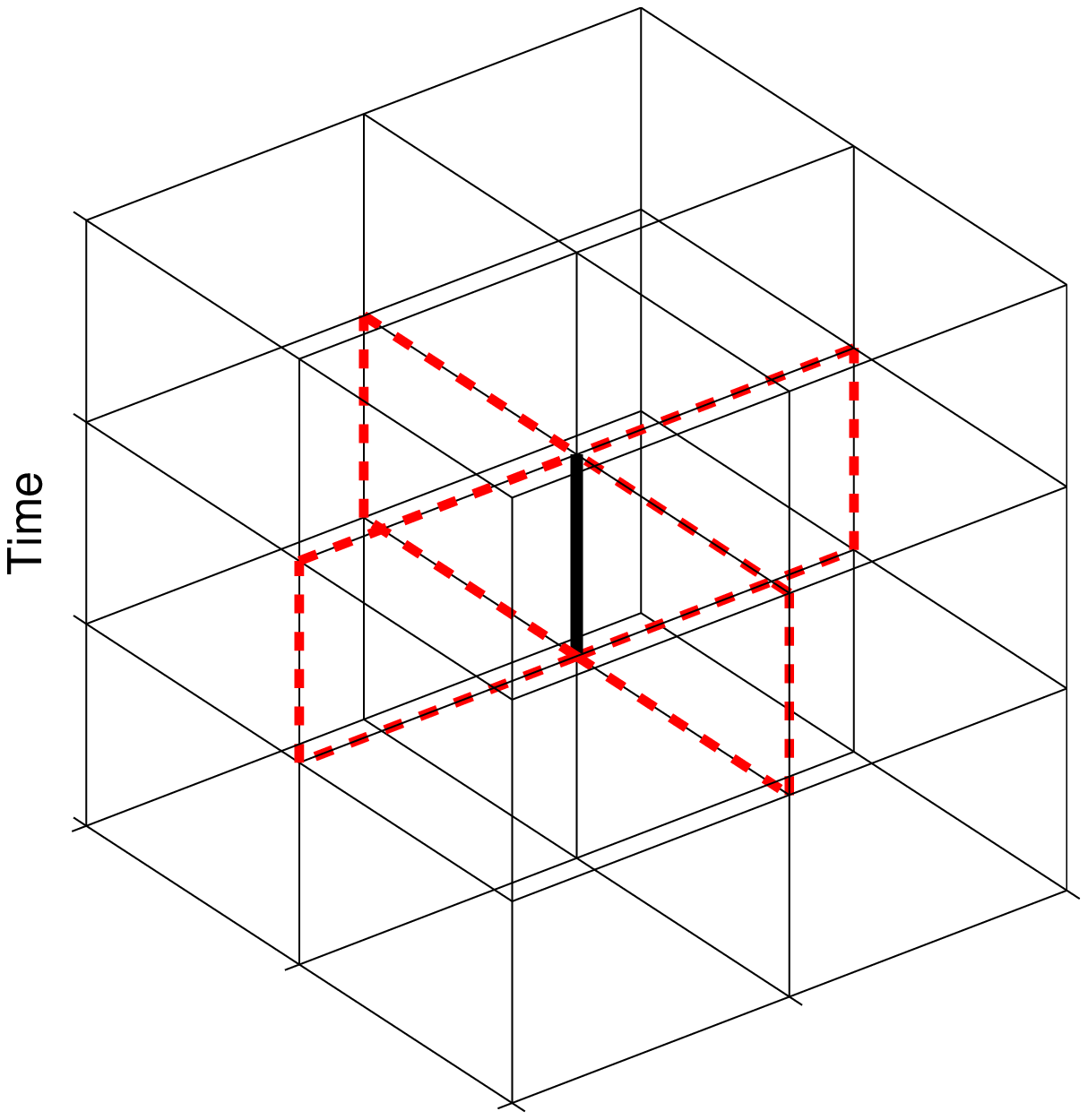}
\includegraphics[width=0.475\textwidth]{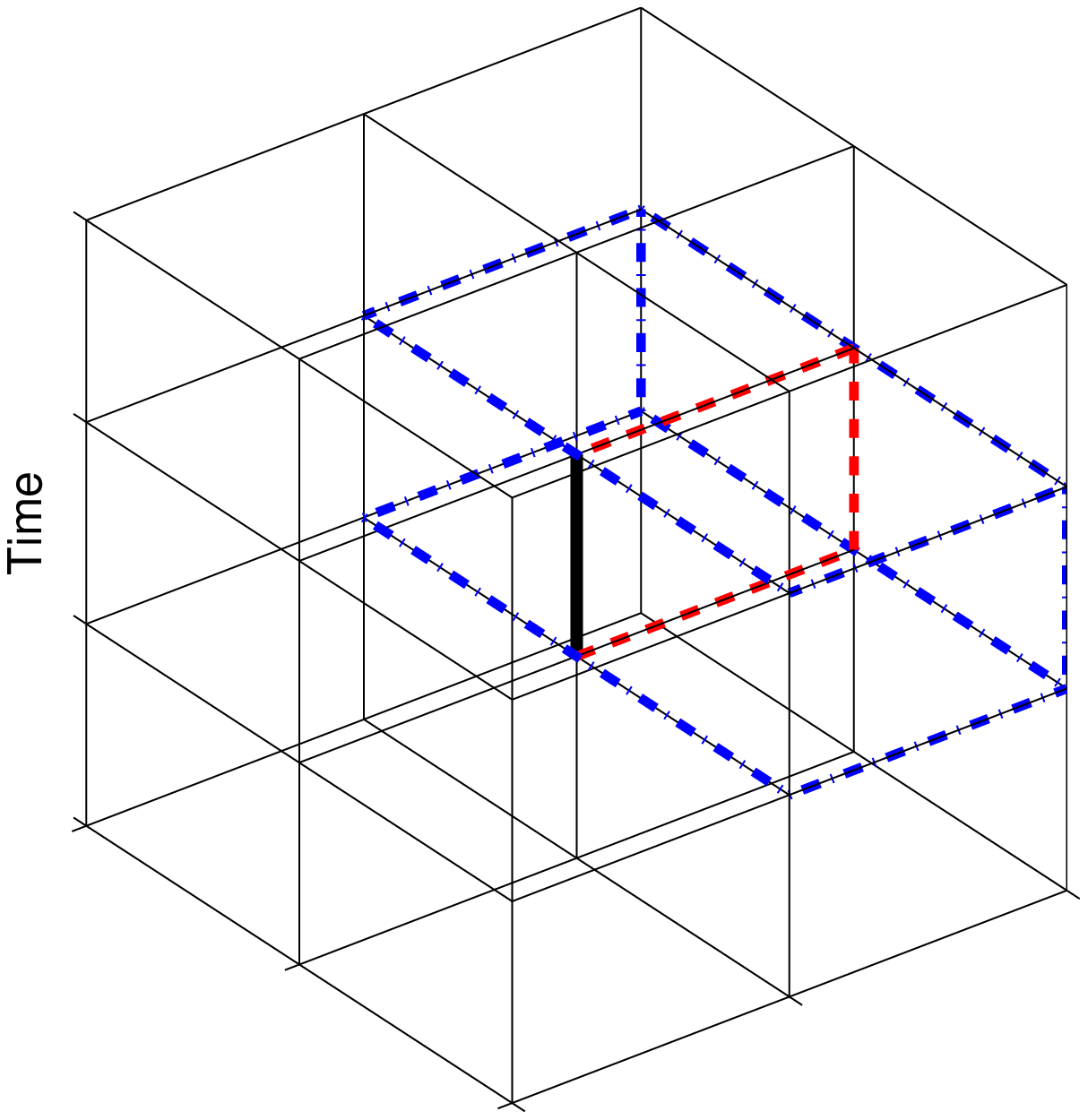}
\caption{(Color online) APE smearing in the time direction (on the left) and hypercubic
blocking (on the right).}
\label{fig:smear}
\end{figure}

We introduce two types of smearing in the time direction to get a better noise to
signal ratio. The first type
is APE type smearing, where the original links in the time direction are replaced
by a sum over the six staples that extend one lattice spacing in the spatial directions
(in Fig.~\ref{fig:smear} on the left). This smearing is called here ``sum6'' for short.
We use the notation ``plain'' to refer to the original Eichten--Hill point static source
construction.

\begin{figure}
\centering
\includegraphics*[angle=-90,width=0.85\textwidth]{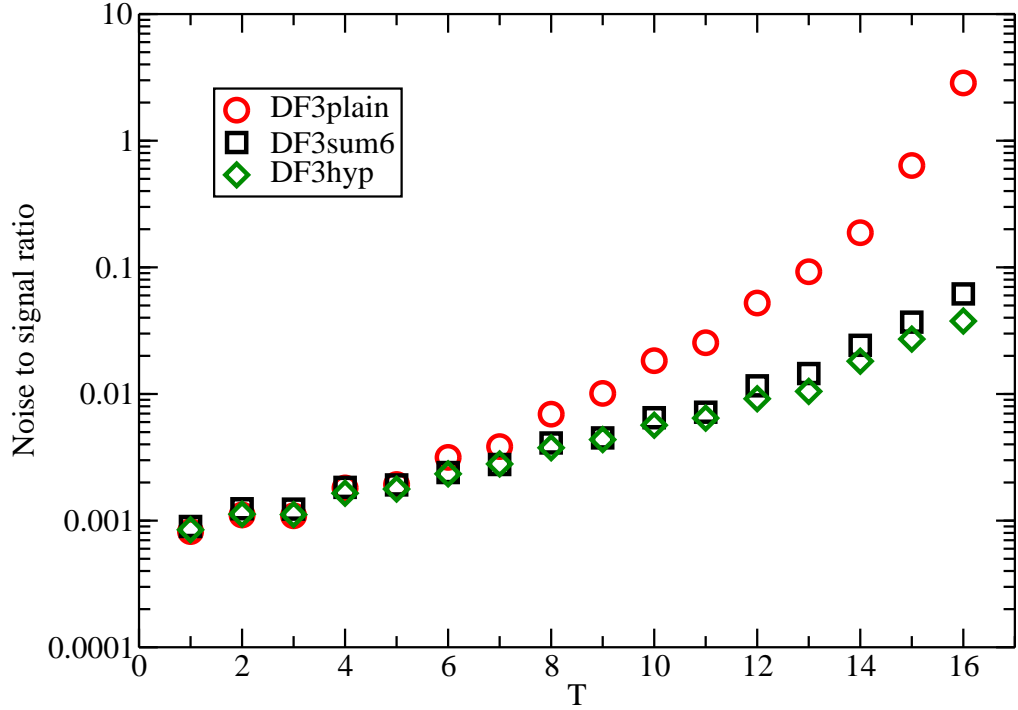}
\caption{(Color online) Noise (standard deviation) to signal ratio: S-wave
2-point correlation function $C_2$ for the lattice ``DF3''.
Note the logarithmic scale.}
\label{fig:rns_S}
\end{figure}

\begin{figure}
\centering
\includegraphics*[angle=-90,width=0.85\textwidth]{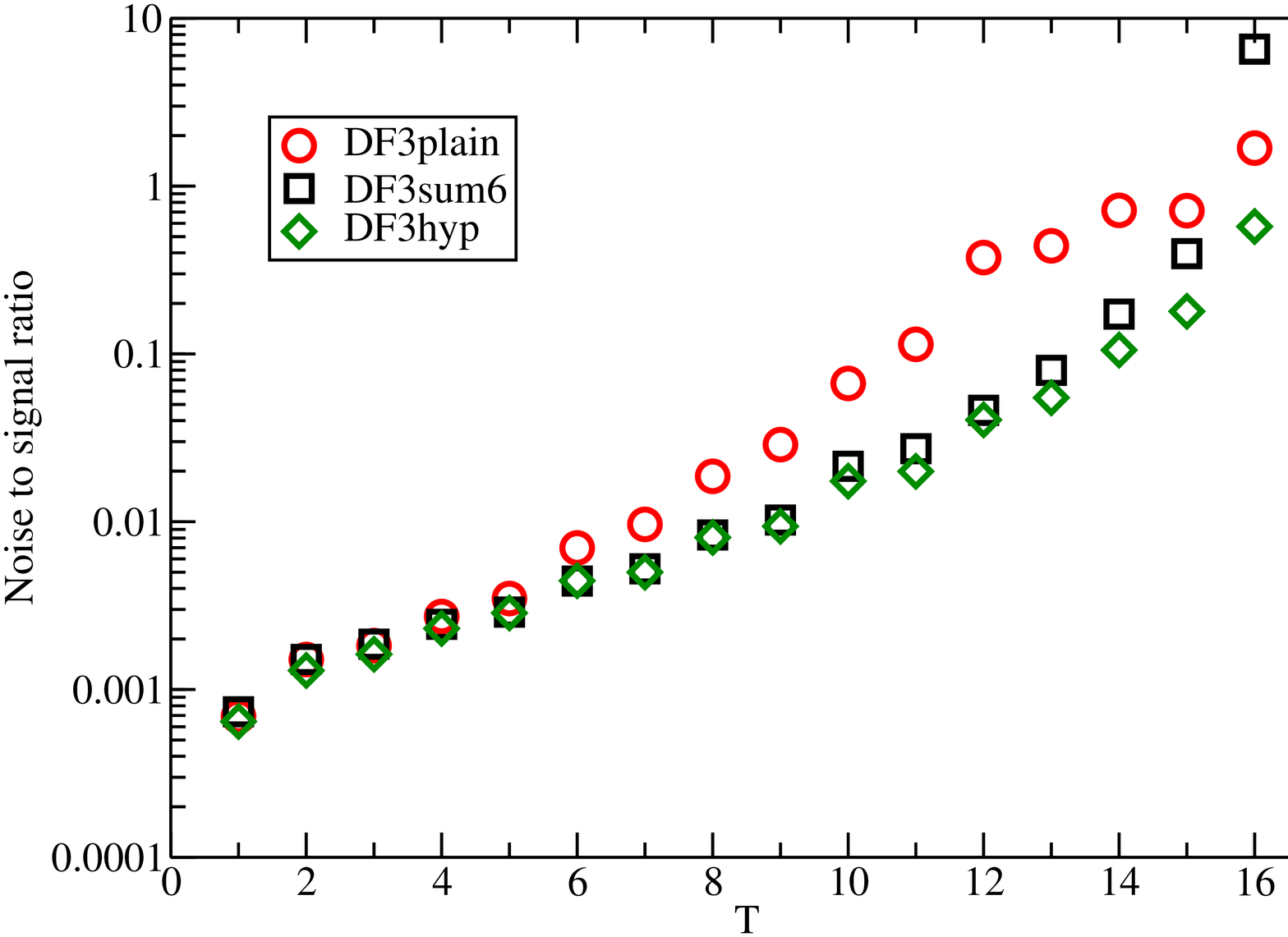}
\caption{(Color online) Noise (standard deviation) to signal ratio
as in Fig.~\ref{fig:rns_S} but for the P$_-$ case.}
\label{fig:rns_Pm}
\end{figure}

\begin{figure}
\centering
\includegraphics*[angle=-90,width=0.85\textwidth]{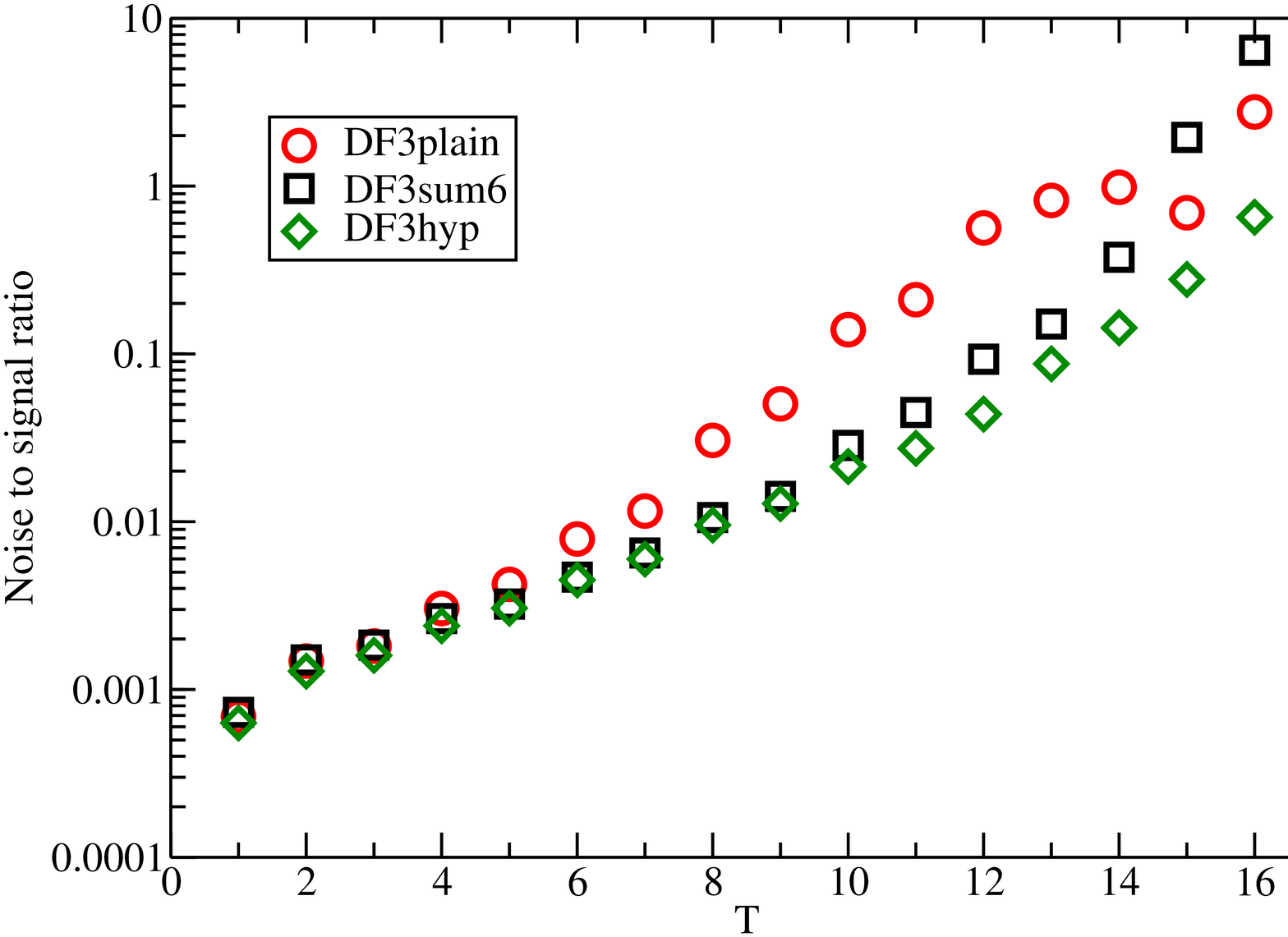}
\caption{(Color online) Noise (standard deviation) to signal ratio for
the P$_+$ case.}
\label{fig:rns_Pp}
\end{figure}

\begin{figure}
\centering
\includegraphics*[angle=-90,width=0.85\textwidth]{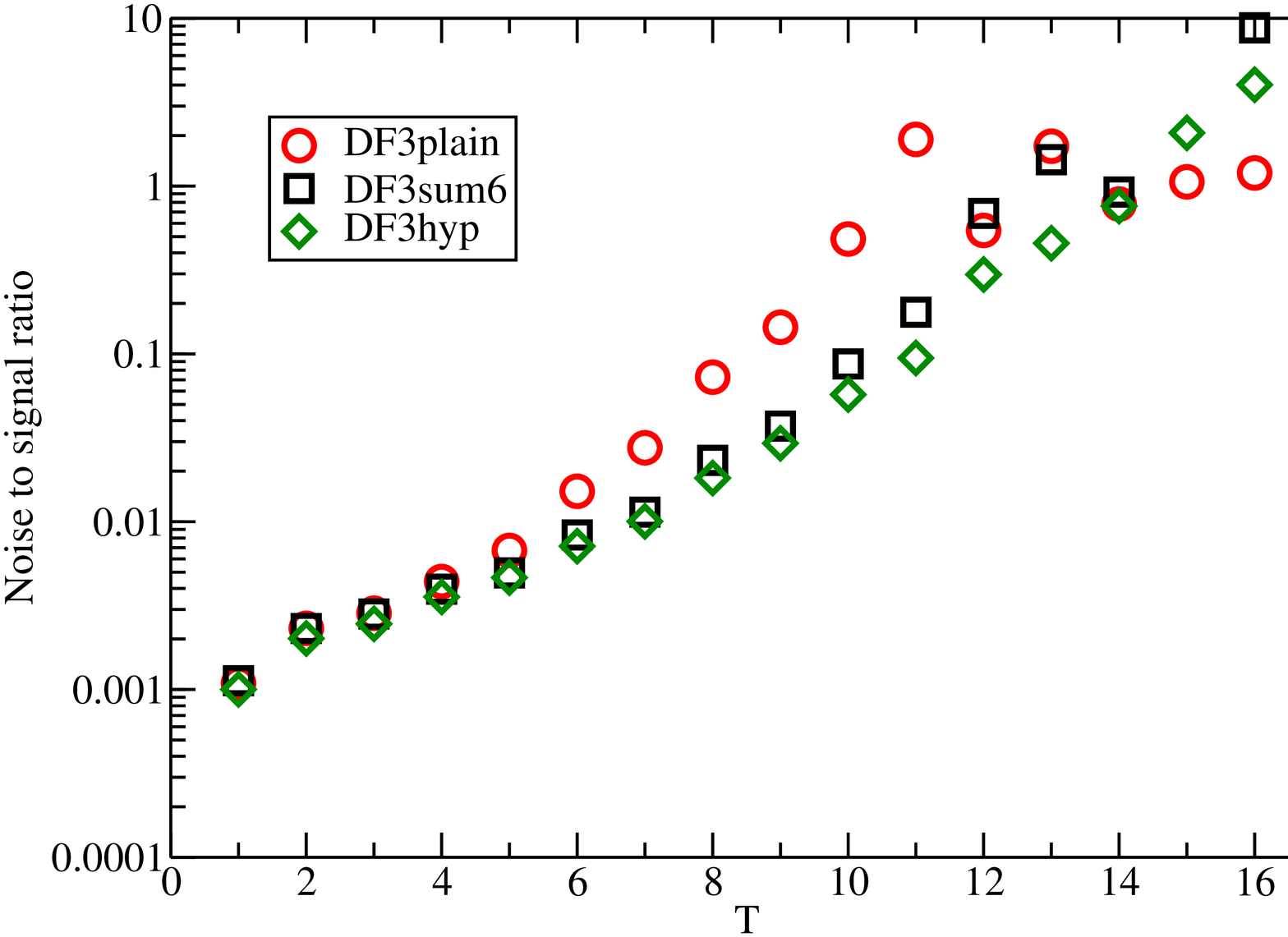}
\caption{(Color online) Noise (standard deviation) to signal ratio for
the D$_-$ case.}
\label{fig:rns_Dm}
\end{figure}

\begin{figure}
\centering
\includegraphics*[angle=-90,width=0.85\textwidth]{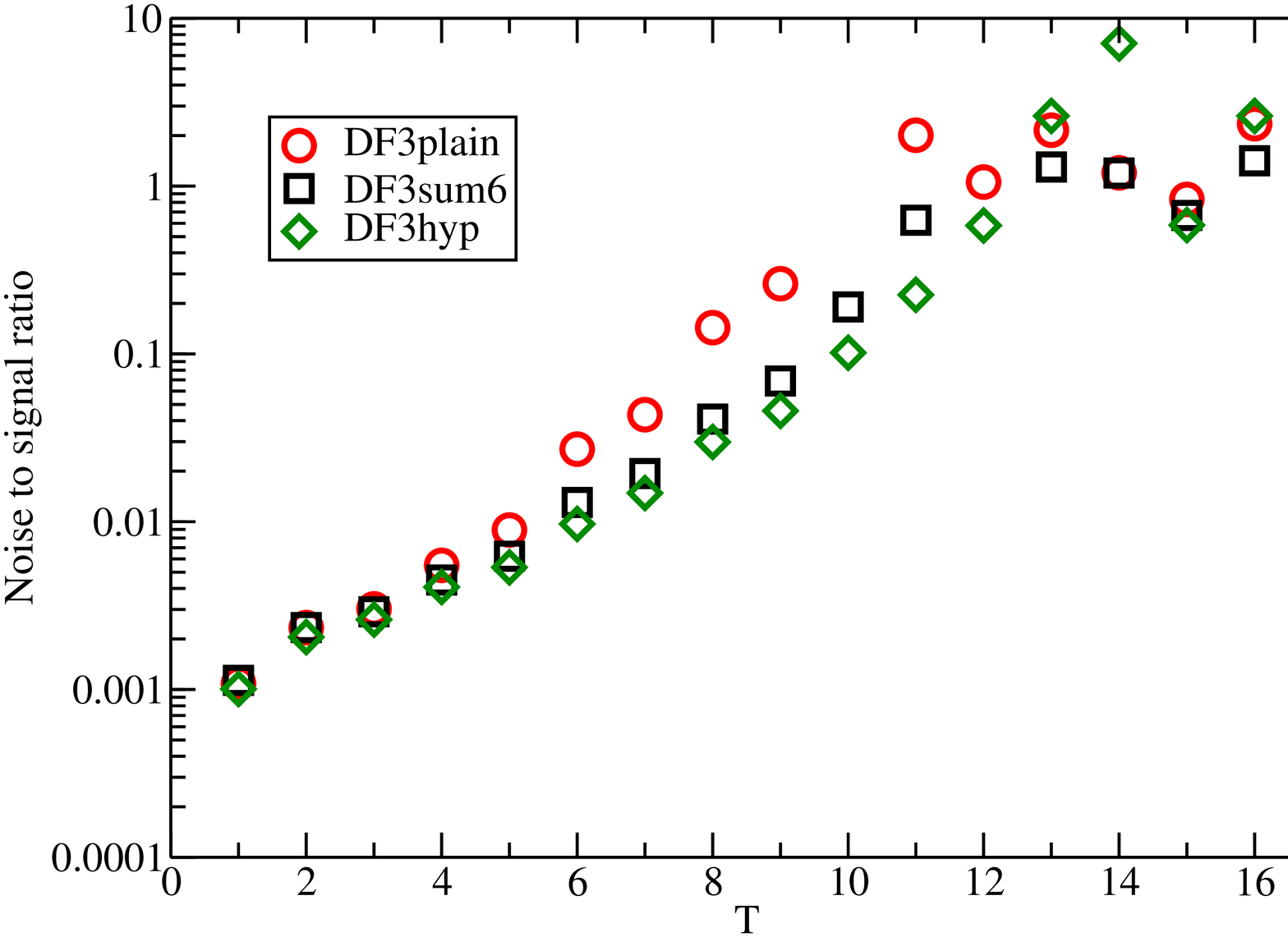}
\caption{(Color online) Noise (standard deviation) to signal ratio for
the D$_+$ case.}
\label{fig:rns_Dp}
\end{figure}

To smear the heavy quark even more we then use hypercubic blocking (first introduced
by Hasenfratz and Knechtli in~\cite{Hasenfratz}), again only
for the links in the time direction (in Fig.~\ref{fig:smear} on the right). Now the
staples (the red dashed lines in Fig.~\ref{fig:smear}) are not constructed
from the original, single links, but from staples (the blue
dash-dotted lines in Fig.~\ref{fig:smear}). In more
detail, we first construct the links
\begin{equation}
\bar{V}_{i,\,\mu\, ;\,\nu \,\rho}=\textrm{Proj}_{\textrm{SU}(3)}\bigg [
(1-\alpha_3)U_{i,\,\mu}+\frac{\alpha_3}{2}\sum_{\pm\eta\,\neq\,\rho ,\,\nu ,\,\mu}
U_{i,\,\eta}U_{i+\hat{\eta} ,\,\mu}U^\dag_{i+\hat{\mu} ,\,\eta} \bigg ],
\end{equation}
where $U_{i,\,\mu}$ is the original thin link at location $i$ and direction $\mu$.
Note that there are no staples in directions $\nu$ or $\rho$. We then construct
``fat'' links
\begin{equation}
\tilde{V}_{i,\,\mu\, ;\,\nu}=\textrm{Proj}_{\textrm{SU}(3)}\bigg [
(1-\alpha_2)U_{i,\mu}+\frac{\alpha_2}{4}\sum_{\pm\rho\,\neq\,\nu ,\,\mu}
\bar{V}_{i,\,\rho\, ;\,\nu \,\mu}\bar{V}_{i+\hat{\rho} ,\,\mu\, ;\,\rho\, \nu}
\bar{V}^\dag_{i+\hat{\mu},\,\rho\, ;\,\nu\, \mu} \bigg ],
\end{equation}
where index $\nu$ indicates that the link is not decorated with staples in that
direction. The last step is
\begin{equation}
V_{i,\,\mu}=\textrm{Proj}_{\textrm{SU}(3)}\bigg [
(1-\alpha_1)U_{i,\,\mu}+\frac{\alpha_1}{6}\sum_{\pm\nu\,\neq\,\mu}
\tilde{V}_{i,\,\nu\, ;\,\mu}\tilde{V}_{i+\hat{\nu} ,\,\mu\, ;\,\nu}
\tilde{V}^\dag_{i+\hat{\mu} ,\,\nu\, ;\,\mu} \bigg ],
\end{equation}
where the ``fat'' links are again used to construct the new links.
The values $\alpha_3=0.5$, $\alpha_2=1$ and $\alpha_1=1$ are used in this study,
because this choice was found to be very good in reducing the noise to
signal ratio in~\cite{Sommer}. Note that $\alpha_3=1.0$, $\alpha_2=0$ and $\alpha_1=0$
would give the ``sum6'' smearing.
Hypercubic blocking takes into account the links within a ``hypercube'' (the edges
of the ``cube'' are 2$a$ in spatial directions but only one lattice spacing in the
time direction). This smearing is called here ``hyp'' for short. The ``plain''
configurations do not have smearing in the time direction.
Smearing the heavy quark improves the noise to signal ratio, which can be seen in
Figs.~\ref{fig:rns_S}--\ref{fig:rns_Dp}. The figures show the standard deviation
to signal ratio for the largest component of $C_2$, which is F2F2, for 160 lattice
configurations (lattice ``DF3''). In all cases the ``plain'' signal is clearly
inferior to the ``sum6'' and ``hyp'' signals,
whereas the ``hyp'' signal is also better than the ``sum6'' signal. This
latter difference wouldbe more apparent in a non-logarithmic scale.
Lattices ``DF4'' and ``DF5'' show similar trends as the ``DF3'' lattice.

\section{Energy spectrum}

\begin{figure}
\centering
\includegraphics*[angle=-90,width=0.9\textwidth]{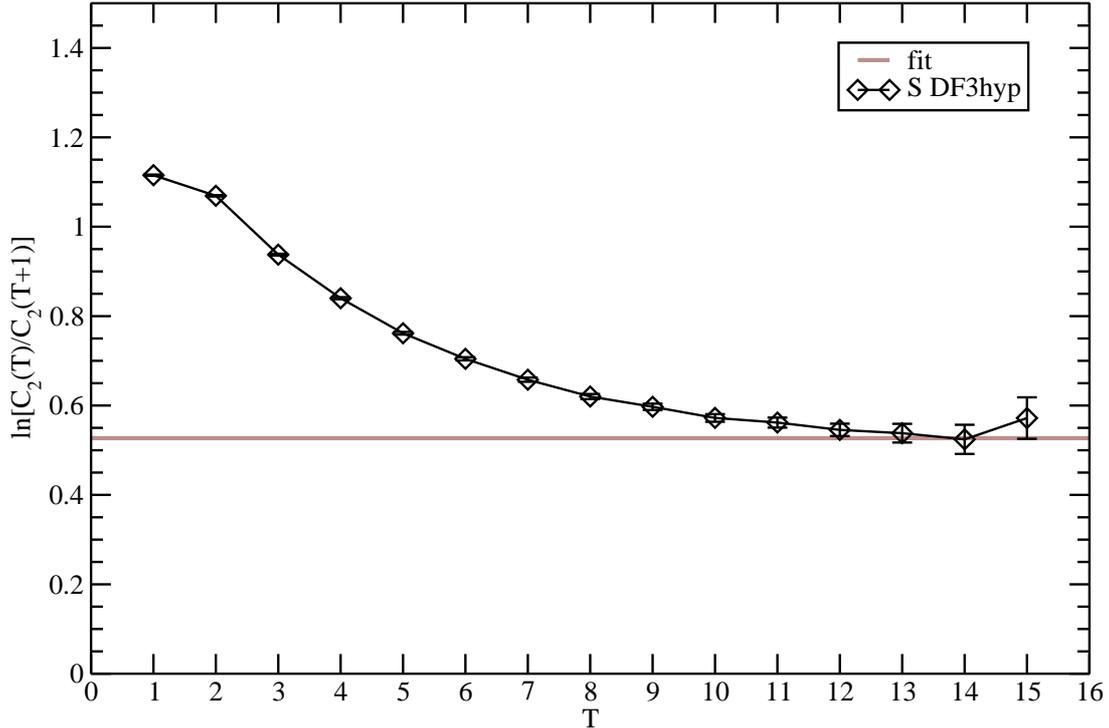}
\caption{(Color online) Effective mass plot for the S-wave F2F2 correlations.
The line labelled ``fit'' (here and
in the other effective mass plots) shows the lowest energy obtained from
the fit in Eq.~\protect\ref{C2fit} for the lattice ``DF3hyp''. Only 2 paths are used
in the fit shown here to make comparisons easier, although the best fit for the
S-wave state uses all 5 paths. The thickness of the line indicates the error. As
expected from a variational argument, this fit to all data naturally gives a somewhat
smaller mass than a fit to F2F2 alone.}
\label{Fig:EffMassS}
\end{figure}

\begin{figure}
\centering
\includegraphics*[angle=-90,width=0.9\textwidth]{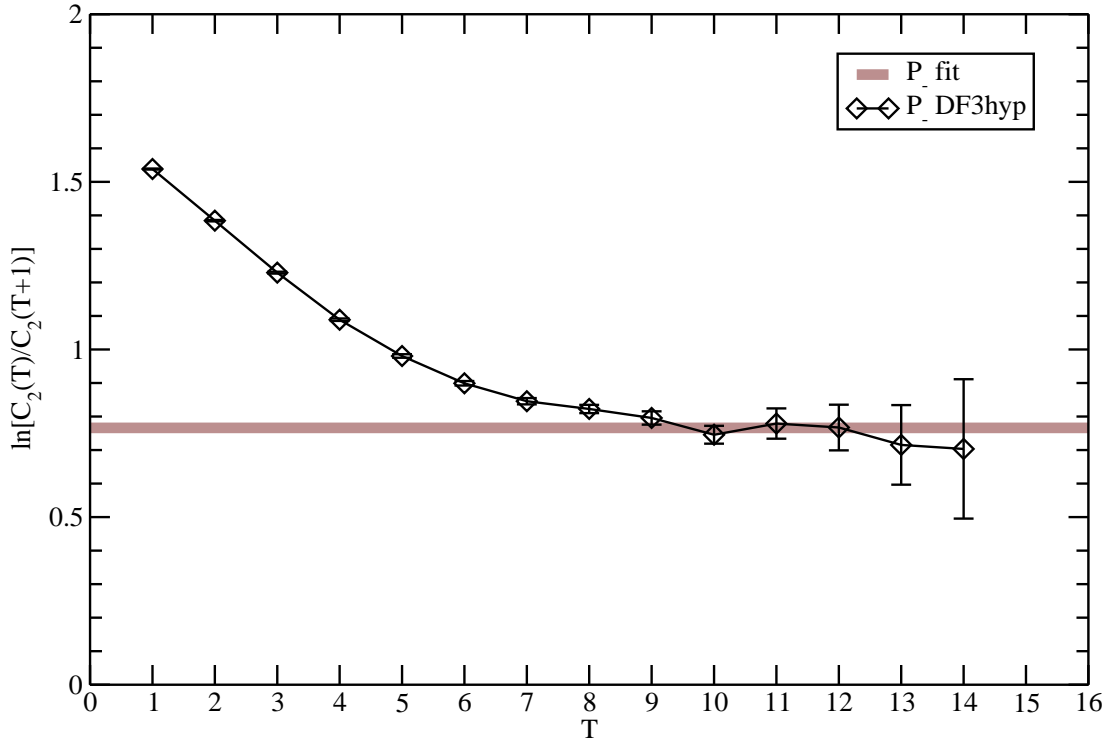}
\caption{(Color online) Effective mass plot for the P$_-$ F2F2 correlations.
Only 2 paths are used in the fit shown here to make comparisons easier, although
the best fit for this state uses all 5 paths. Other details as in
Fig.~\ref{Fig:EffMassS}.}
\label{Fig:EffMassPm}
\end{figure}

\begin{figure}
\centering
\includegraphics*[angle=-90,width=0.9\textwidth]{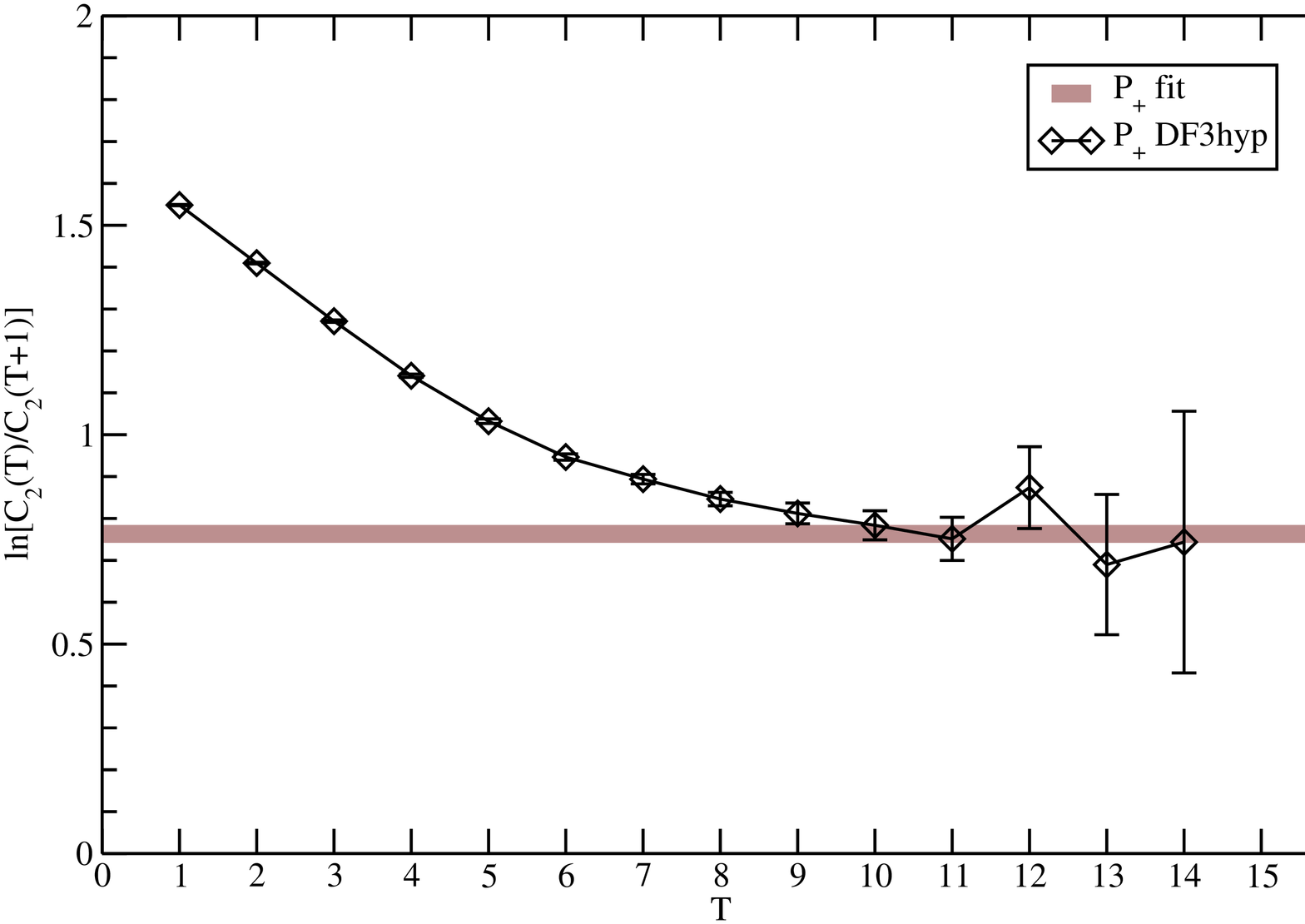}
\caption{(Color online) Effective mass plot for the P$_+$ F2F2 correlations.
Other details as in Fig.~\ref{Fig:EffMassS}.}
\label{Fig:EffMassPp}
\end{figure}

\begin{figure}
\centering
\includegraphics*[angle=-90,width=0.9\textwidth]{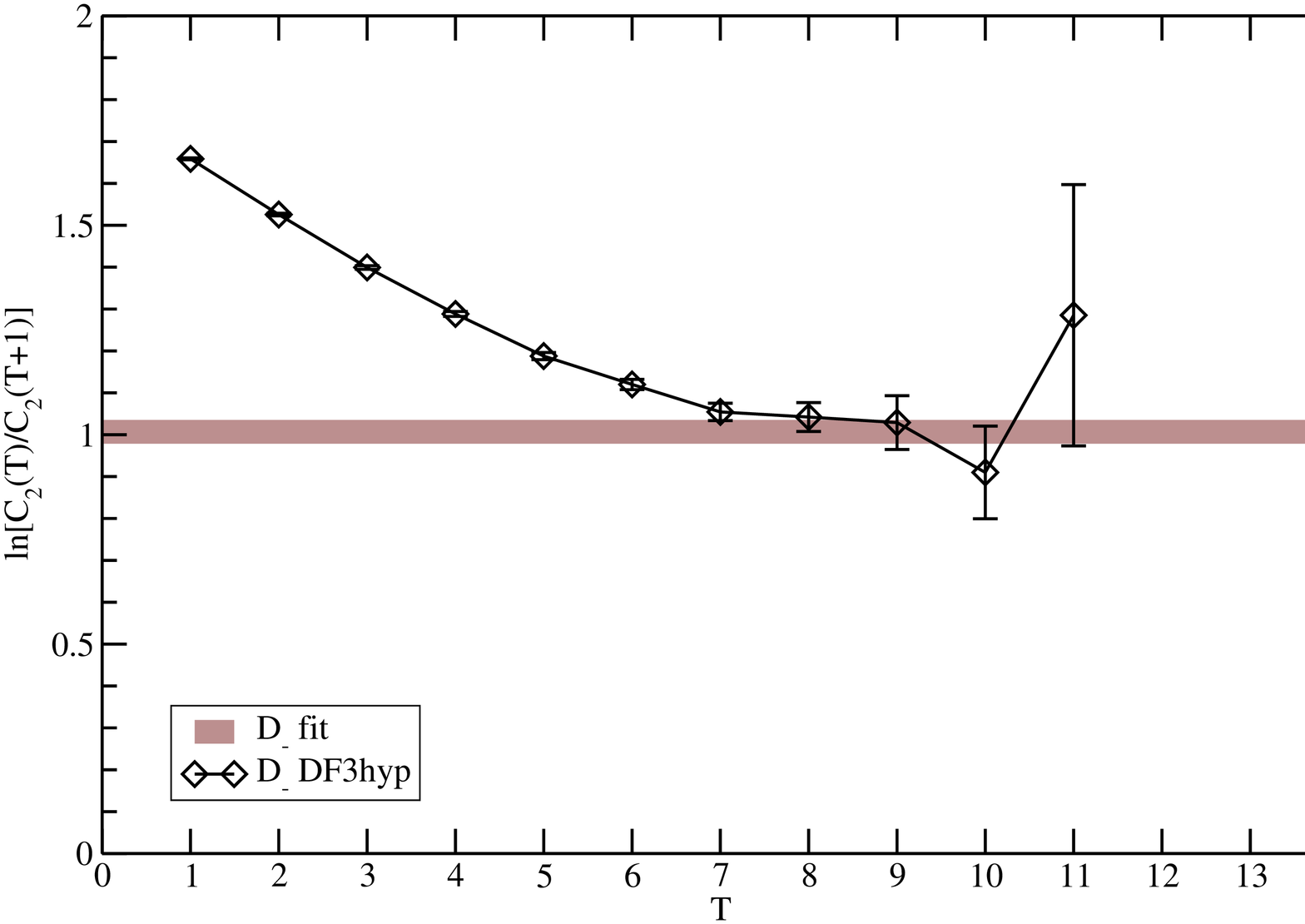}
\caption{(Color online) Effective mass plot for the D$_-$ F2F2 correlations.
Other details as in Fig.~\ref{Fig:EffMassS}.}
\label{Fig:EffMassDm}
\end{figure}

\begin{figure}
\centering
\includegraphics*[angle=-90,width=0.9\textwidth]{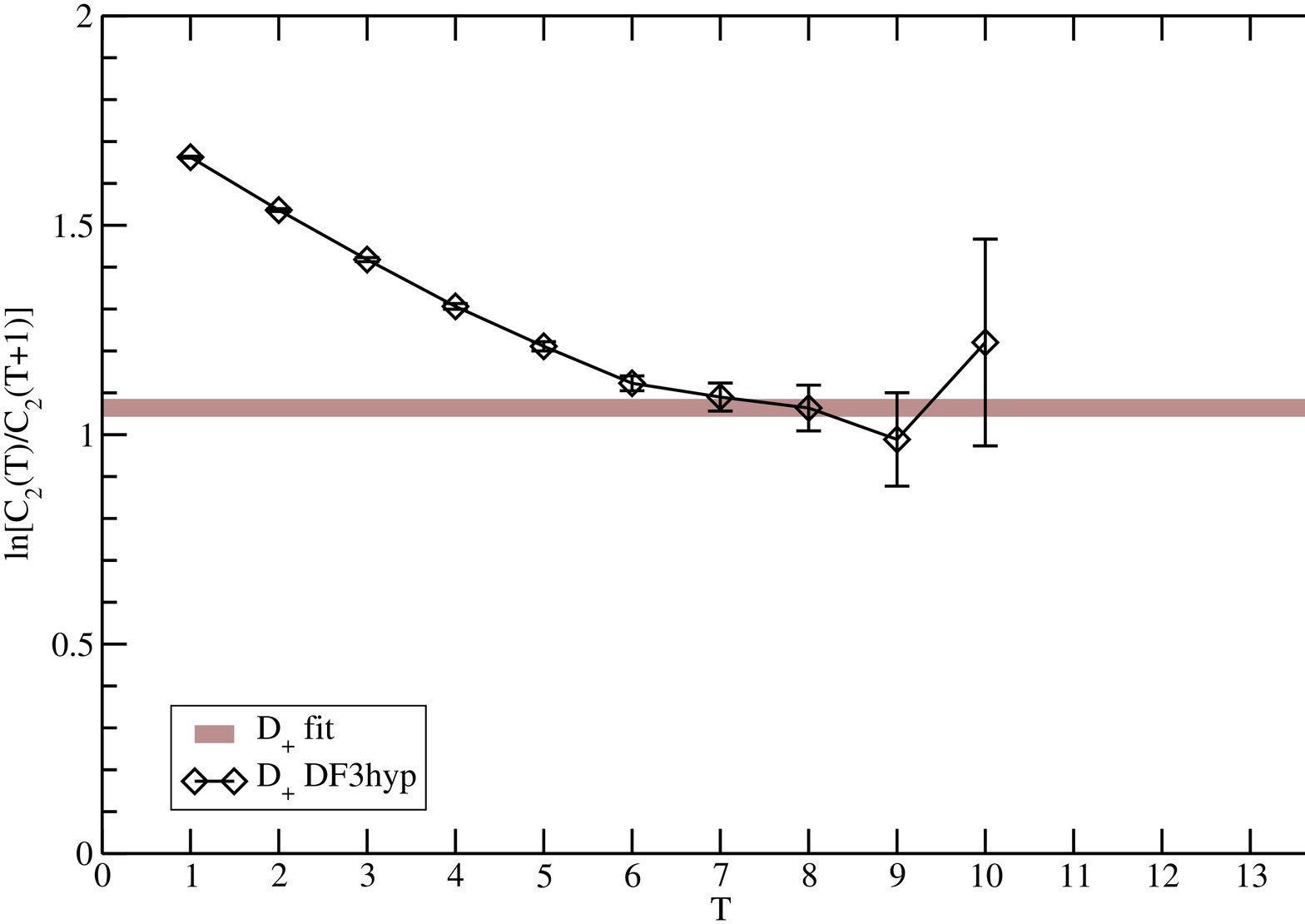}
\caption{(Color online) Effective mass plot for the D$_+$ F2F2 correlations.
Other details as in Fig.~\ref{Fig:EffMassS}.}
\label{Fig:EffMassDp}
\end{figure}

\begin{figure}
\centering
\includegraphics*[angle=-90,width=0.9\textwidth]{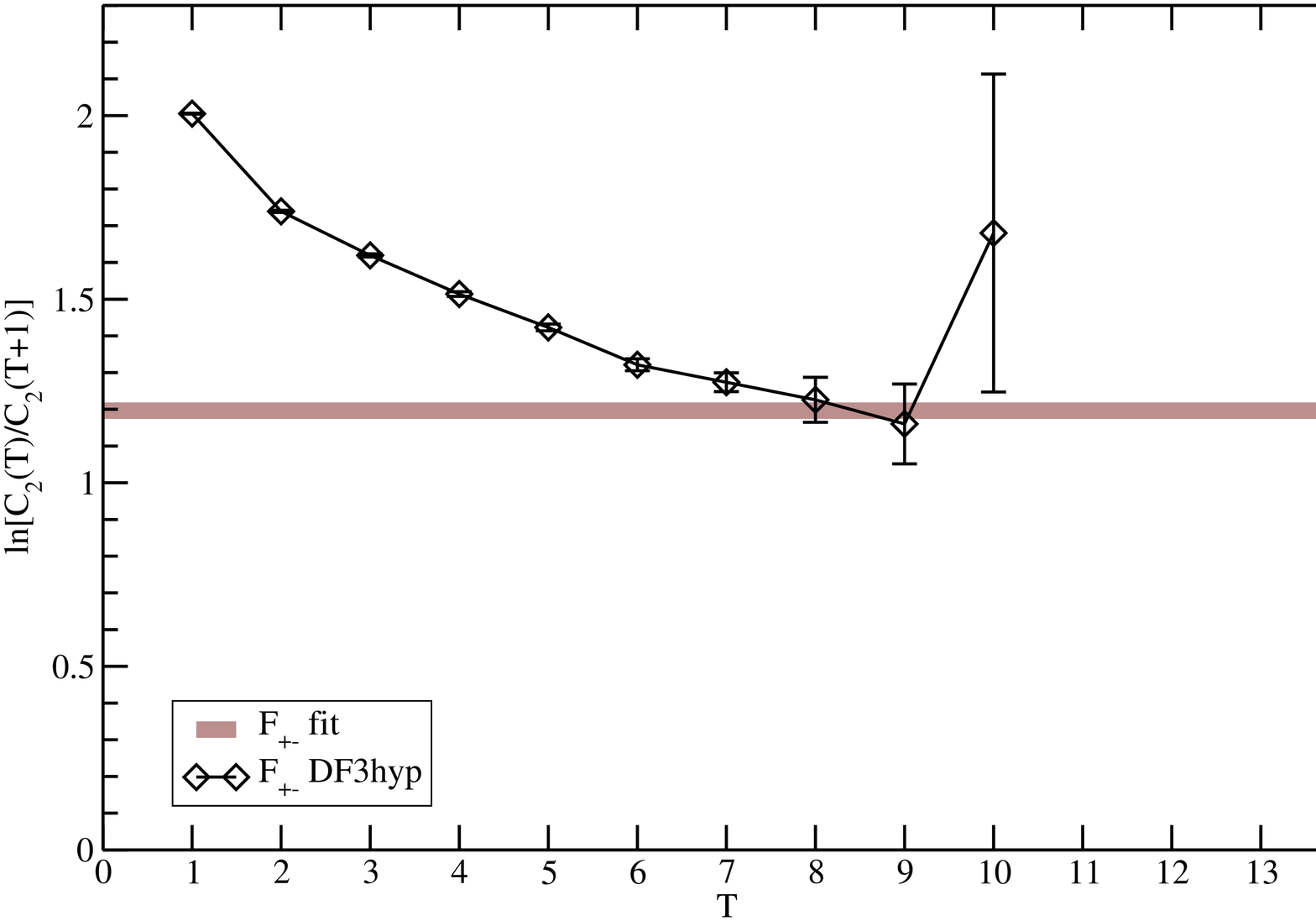}
\caption{(Color online) Effective mass plot for the F$_{+-}$ F2F2 correlations.
Other details as in Fig.~\ref{Fig:EffMassS}.}
\label{Fig:EffMassF}
\end{figure}

\begin{table}
\centering
 \begin{tabular}{|c|ccc|cccccc|c|}
 \hline
 L$_{\pm}$& $m_1$  & $m_2$  & $m_3$  & $a_{1, \textrm{F1}}$ & $a_{1, \textrm{F2}}$ & $a_{2, \textrm{F1}}$ & $a_{2, \textrm{F2}}$ & $a_{3, \textrm{F1}}$ & $a_{3, \textrm{F2}}$ & $\frac{\chi^2}{\textrm{d.o.f.}}$ \\
 \hline
 S        & 0.527(5)  & 0.98(2)  & 1.40(2)          & 0.297(11) & 0.44(2)  & 1.24(5)  & 1.56(4)  & 1.81(5)  & 0.03(11) & 21/24 \\
 P$_-$    & 0.766(14) & 1.29(3)  & 1.52(3)          & 0.65(5)   & 0.93(7)  & 3.4(3)   & 2.2(2)   & -1.5(6)  & 3.4(6)   & 19/24 \\
 P$_+$    & 0.76(2)   & 1.28(3)  & 1.46(2)          & 0.62(7)   & 0.88(10) & 3.6(4)   & 3.14(15) & -2.8(5)  & 2.6(5)   & 11/24 \\
 D$_{+-}$ & 1.10(8)   & 1.46(5)  & 1.66(6)$^\ast$   & 1.0(3)    & 2.0(5)   & 5.3(8)   & 1.8(9)   & -0.6(17) & 5.1(13)  & 22/24 \\
 D$_-$    & 1.01(3)   & 1.52(2)  & 1.67(4)$^\ast$   & 1.10(14)  & 2.0(2)   & 7.2(5)   & 1.5(12)  & -1.4(15) & 7.4(8)   & 36/27 \\
 D$_+$    & 1.06(2)   & 1.558(8) & 1.80(2)$^\ast$   & 1.14(9)   & 2.09(11) & 6.52(6)  & -0.5(5)0 & 1.3(7)   & 7.6(2)   & 39/27 \\
 F$_{+-}$ & 1.20(2)   & 1.658(5) & 1.892(13)$^\ast$ & 0.77(8)   & 1.88(10) & 5.30(13) & -1.1(5)  & 1.3(7)   & 6.0(3)   & 37/27 \\
 \hline
\end{tabular}
\caption{Two-point correlation function fits (equation \ref{C2fit}) for ``DF3hyp''. 2 path fit
results are shown for all states to make comparisons easier, even though our best fits for the S and
P$_-$ states are 5 path fits. In some cases (entries marked with an asterisk) Bayesian ideas are
used (see section~\ref{bayes}). The errors on the parameters were obtained by bootstrapping
the lattice configurations and repeating the fit 100 times. The $m_i$ are in lattice units.
Note that the $\frac{\chi^2}{\textrm{d.o.f.}}$ is larger for the D-wave and F-wave states than
for the S- and P-wave states.}
\label{C2FitTable}
\end{table}

The energies are obtained from the fit in equation~\ref{C2fit} --- see Table~\ref{C2FitTable}
for the results for the lattice ``DF3hyp''. The $m_i$ are in lattice units. However, due to the
presence of an unknown (but L$_{\pm}$ and $i$ independent) self energy in each $m_i$, only the
differences $m_i(\textrm{L}_{\pm})-m_1(\textrm{S})$ are relevant. 
The ground state energy from the 2-path fit for a
given state is compared with the effective mass in Figs.~\ref{Fig:EffMassS}-\ref{Fig:EffMassDp}.
To illustrate how plateaux develop with $T$, $\ln{[C_2(T)/C_2(T+1)]}$ is shown for the
largest component, F2F2, for the same lattice ``DF3hyp''. It is seen that for all states we
get a plateau that agrees nicely with the fit result that uses all components, F1F1, F1F2
and F2F2. The errors are large for large $T$ values, and the data points for the highest $T$ values
are not shown if the errors render them insignificant. In the fits this is under better
control, because we fit $C_2(T)$ and not the ratio, and all fuzzing combinations are used
(i.e. more data are used). We can thus use data up to $T=15$ in the fits. The fit shown in these
figures is only to the 2 path data, to make comparisons easier. When extracting the energy
of a state all 5 paths are used for S and P$_-$. For other states only 2 path data are available.

\begin{figure}
\centering
\includegraphics[height=0.60\textwidth, angle=-90]{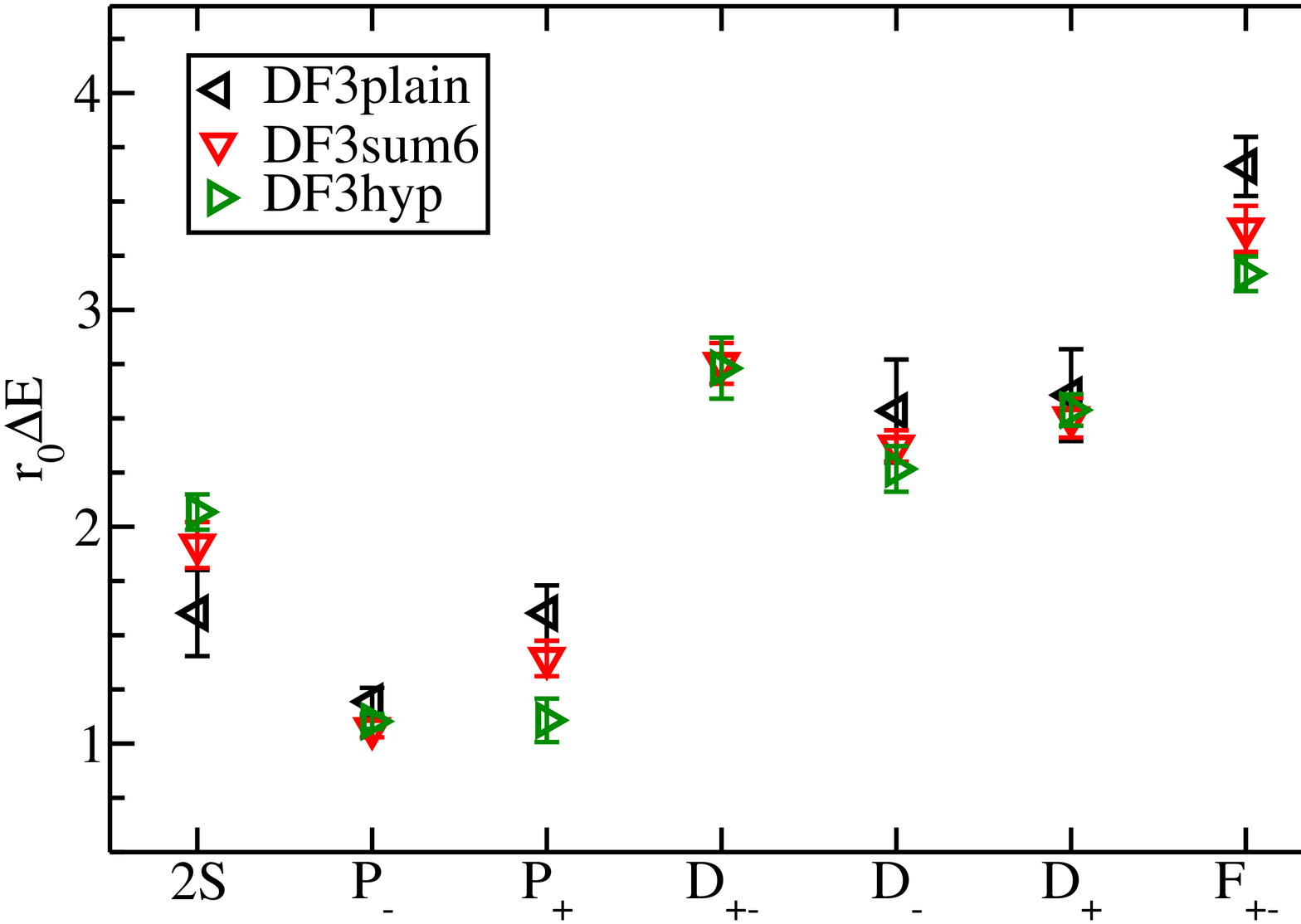}
\caption{(Color online) Energy spectrum of the heavy-light meson using lattice ``DF3''
in units of $r_0$. Here L$_+$($_-$) means that the light quark spin couples to angular
momentum L giving the total $j=\textrm{L}\pm 1/2$. The 2S is the first radially
excited L$=0$ state. The D$_{+-}$ is a mixture of the D$_-$ and D$_+$ states, and
likewise for the F$_{+-}$.  Energies are given with respect to the S-wave
ground state (1S). Here $r_0/a=4.754(40)^{+2}_{-90}$ (from~\cite{Allton}).
The error bars shown here contain only the statistical errors on the lattice
energy fits.
}
\label{fig:espectr}
\end{figure}

\begin{figure}
\centering
\includegraphics[height=0.60\textwidth, angle=-90]{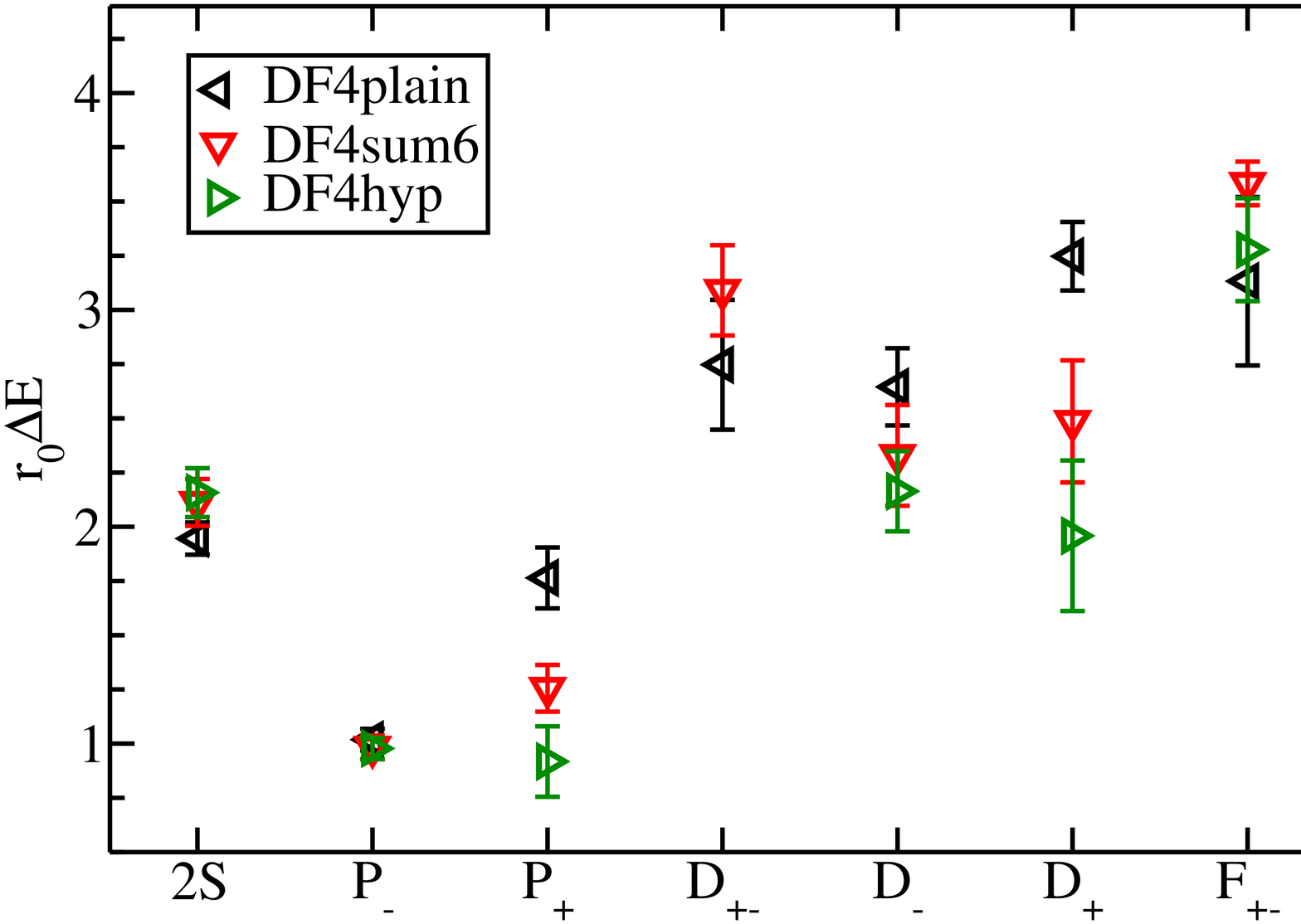}
\caption{(Color online) Energy spectrum of the heavy-light meson using lattice ``DF4''.
See Fig.~\ref{fig:espectr} for details. Here $r_0/a=5.041(40)^{+0}_{-10}$
(from~\cite{Allton}). The error bars shown here contain only the statistical errors
on the lattice energy fits.
}
\label{fig:espectr2}
\end{figure}

\begin{figure}
\centering
\includegraphics[height=0.60\textwidth, angle=-90]{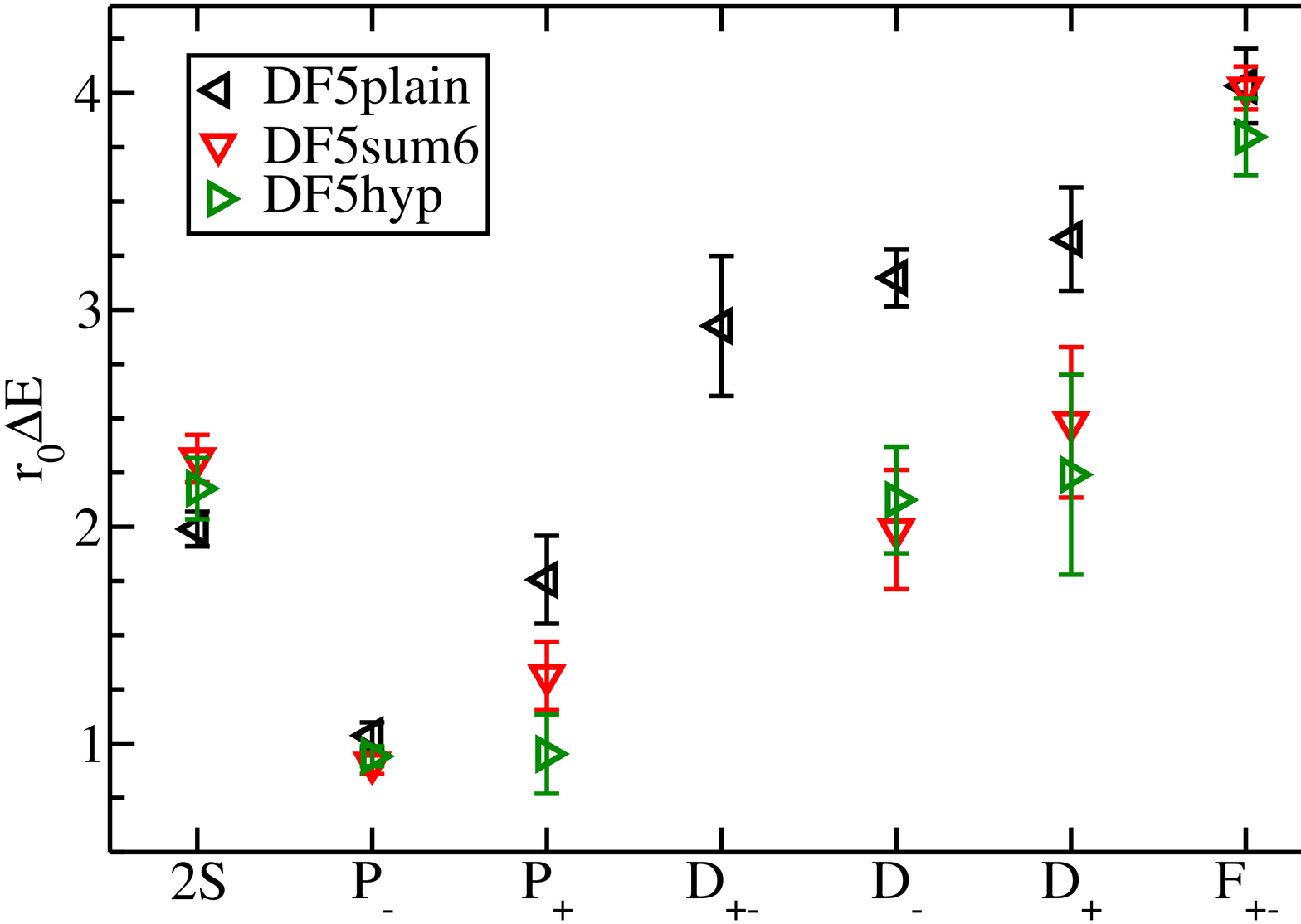}
\caption{(Color online) Energy spectrum of the heavy-light meson using lattice ``DF5''.
See Fig.~\ref{fig:espectr} for details. Here $r_0/a=5.32(5)$ (from~\cite{Allton}).
The error bars shown here contain only the statistical errors from the lattice energy
fits.
}
\label{fig:espectr3}
\end{figure}

The resulting energy spectra from the fit (Eq.~\ref{C2fit}) for different
lattices are shown in Figs.~\ref{fig:espectr}--\ref{fig:espectr3}
--- see also Tables~\ref{EnergyTableDF3}--\ref{EnergyTableDF5}.
With the lattice ``DF3'', for most states, using different smearing for the
heavy quark does not seem to change significantly the energy differences with
respect to the 1S energy --- the exceptions being the P$_+$ and excited D$_{+-}$
states. Different smearings should only give the same results in the continuum
limit, so it is understandable that at a fixed lattice spacing the results may
differ. Unfortunately, all our lattices have approximately the same lattice spacing
(about $a=0.1$~fm, see Table~\ref{LatParams}), and we can not go to the continuum limit
properly. However, we can use the results from different smearings to give a rough
estimate of the systematic error. Because studying the noise to signal ratio
(Figs.~\ref{fig:rns_S}--\ref{fig:rns_Dp}) shows that the ``plain'' configurations
are clearly inferior to the configurations that are smeared in the time direction,
we use ``hyp'' and ``sum6'' configurations to get our main results.
The reason for quoting energies in units of $r_0$ is to avoid the 5\% uncertainty
in $r_0=0.525(25)$~fm. The uncertainties in $r_0/a$ are much smaller. To get
energies in GeV then requires an additional factor of 0.38(2).

The energy of the D$_{+-}$ state had been expected
to be near the spin average of the D$_-$ and D$_+$ energies, but it turns out to be
a poor estimate of this average. Therefore, it is not clear to what extent
the F$_{+-}$ energy is near the spin average of the two F-wave states, as was
originally hoped. The F- and D-wave results for different
smearings for the lattices ``DF4'' and ``DF5'' are somewhat more scattered than
for the ``DF3'' lattice (i.e. the systematic errors are larger), but otherwise
the same features are seen for all three light quark masses. We are most
interested in the ``DF3'' lattice results, because that is closest to the $B_s$
meson (the light quark mass on this lattice being close to the \textit{s} quark mass).
One interesting observation is, that the energy spectrum is close to being dependent
on L alone. For example our preferred configurations, ``DF3hyp'', show an approximate
linear rise in excitation energy with L (up to F-wave) as $\sim$0.4L GeV. A similar
linear rise is usually seen in Regge or string models. In contrast, the 2S state is
seen to be almost degenerate with the 1D states, as in a simple harmonic oscillator.
A L(L+1) term can be added to the linear ansatz to get a better fit --- more
precisely, 0.34L+0.04L(L+1) gives a good overall fit to the four energy
differences up to D-waves.

Our earlier results (for the ``plain'' configurations
used in this study as well as for some other unquenched and quenched configurations)
can be found in Refs.~\cite{PRD69,LattConf}. Because dynamical fermions are used in
this study, we can not be absolutely certain that the lattice states are pure
quark--anti-quark states. However, our radial distribution measurements
\cite{radial} support the assumption that the states are ordinary meson states:
the radial distributions of the lowest lying states are not broad, as would be the
case if the states were molecules, and the first radial excitations of S- and P-wave
states have one node at short distances (approximately at 0.30--0.35~fm).

\begin{table}
\centering
 \begin{tabular}{|c|ccc||c|ccc|}
 \hline
 nL$_{\pm}$ &\, DF3plain\, &\, DF3sum6\, &\, DF3hyp\,  & nL$_{\pm}$ &\, DF3plain\, &\, DF3sum6\, &\, DF3hyp\, \\
 \hline
 1S       & 3.55(6) & 2.724(14)& 2.520(10)&  2S       & 5.1(2)  & 4.64(10) & 4.59(8)  \\
 1P$_-$   & 4.74(3) & 3.79(4)  & 3.62(3)  &  2P$_-$   & 7.13(5) & 5.98(10) & 5.97(10) \\
 1P$_+$   & 5.15(11)& 4.12(8)  & 3.63(10) &  2P$_+$   & 7.80(9) & 6.62(8)  & 6.1(2)   \\
 1D$_{+-}$&  -      & 5.48(9)  & 5.25(14) &  2D$_{+-} $& -       & 6.88(4)  & 7.05(7)  \\
 1D$_-$   & 6.1(2)  & 5.10(7)  & 4.79(13) &  2D$_-$   & 8.44(14)& 7.50(4)  & 7.23(11) \\
 1D$_+$   & 6.2(2)  & 5.23(9)  & 5.06(7)  &  2D$_+$   & 8.38(15)& 7.59(5)  & 7.41(4)  \\
 1F$_{+-}$& 7.21(12)& 6.10(11) & 5.69(8)  &  2F$_{+-}$& 9.16(3) & 8.18(3)  & 7.88(2)  \\
 \hline
\end{tabular}
\caption{Heavy-light meson energies on the lattice in units of $r_0$ for ``DF3''. 
The uncertainty due to the statistical error on $r_0/a$
[$r_0/a=4.754(40)^{+2}_{-90}$, from~\cite{Allton}] is small (less than 1\% ), and is
not taken into account in the error estimates. 
The n denotes the radial excitation and n$-1$ gives the number of nodes
in the wavefunction of the state. The dash means that no reliable fit can be found.
The results on different lattices can not be compared directly (only energy differences
can be compared) due to different self energies. The ``DF3hyp'' results are the same as
$m_1$, $m_2$ in Table~\ref{C2FitTable}, but expressed in different units. Also the
1S, 1P$_-$ are now from 5 path fits.}
\label{EnergyTableDF3}
\end{table}


\begin{table}
\centering
 \begin{tabular}{|c|ccc||c|ccc|}
 \hline
 nL$_{\pm}$ &\, DF4plain\, &\, DF4sum6\, &\, DF4hyp\,  & nL$_{\pm}$ &\, DF4plain\, &\, DF4sum6\, &\, DF4hyp\, \\
 \hline
 1S       & 3.72(4)  & 2.66(2)  & 2.45(2)  &  2S       & 5.66(7)  & 4.77(11) & 4.61(11) \\
 1P$_-$   & 4.73(4)  & 3.64(5)  & 3.43(5)  &  2P$_-$   & 7.38(6)  & 6.03(12) & 5.91(11) \\
 1P$_+$   & 5.48(14) & 3.92(11) & 3.37(16) &  2P$_+$   & 8.27(11) & 6.65(9)  & 6.0(2)   \\
 1D$_{+-}$& 6.5(3)   & 5.8(2)   & -        &  2D$_{+-}$& 8.24(12) & 7.25(9)  & -        \\
 1D$_-$   & 6.4(2)   & 5.0(2)   & 4.6(2)   &  2D$_-$   & 8.91(8)  & 7.6(2)   & 7.2(2)   \\
 1D$_+$   & 6.96(15) & 5.1(3)   & 4.4(3)   &  2D$_+$   & 9.16(8)  & 7.9(2)   & 7.1(2)   \\
 1F$_{+-}$& 6.8(4)   & 6.24(10) & 5.7(2)   &  2F$_{+-}$& 9.51(11) & 8.62(2)  & 8.33(10) \\
 \hline
\end{tabular}
\caption{Heavy-light meson energies on the lattice in units of $r_0$ for ``DF4''. 
The uncertainty due to the statistical error on $r_0/a$
[$r_0/a=5.041(40)^{+0}_{-10}$, from~\cite{Allton}] is small (less than 1\% ), and
is not taken into account in the error estimates. Other comments as in
Table~\ref{EnergyTableDF3}.}
\label{EnergyTableDF4}
\end{table}

\begin{table}
\centering
 \begin{tabular}{|c|ccc||c|ccc|}
 \hline
 nL$_{\pm}$ &\, DF5plain\, &\, DF5sum6\, &\, DF5hyp\,  & nL$_{\pm}$ &\, DF5plain\, &\, DF5sum6\, &\, DF5hyp\, \\
 \hline
 1S       & 3.71(5)  & 2.70(3)  & 2.46(3) &  2S       &  5.70(6)  & 5.01(11) & 4.63(14) \\
 1P$_-$   & 4.75(4)  & 3.61(4)  & 3.40(4) &  2P$_-$   &  7.57(6)  & 6.23(10) & 6.11(10) \\
 1P$_+$   & 5.5(2)   & 4.01(15) & 3.4(2)  &  2P$_+$   &  8.51(12) & 6.96(11) & 6.3(2)   \\
 1D$_{+-}$& 6.6(3)   & -        & -       &  2D$_{+-}$&  8.71(13) & -        & -        \\
 1D$_-$   & 6.86(12) & 4.7(3)   & 4.6(2)  &  2D$_-$   &  9.51(5)  & 7.5(2)   & 7.4(2)   \\
 1D$_+$   & 7.0(2)   & 5.2(3)   & 4.7(5)  &  2D$_+$   &  9.61(8)  & 8.2(2)   & 7.6(3)   \\
 1F$_{+-}$& 7.7(2)   & 6.72(10) & 6.3(2)  &  2F$_{+-}$& 10.05(6)  & 9.01(3)  & 8.6(3)   \\
 \hline
\end{tabular}
\caption{Heavy-light meson energies on the lattice in units of $r_0$ for ``DF5''. 
The uncertainty due to the statistical error on $r_0/a$
[$r_0/a=5.32(5)$, from~\cite{Allton}] is small (less than 1\% ), and is not
taken into account in the error estimates.
Other comments as in Table~\ref{EnergyTableDF3}.}
\label{EnergyTableDF5}
\end{table}

\begin{figure}
\centering
\includegraphics[angle=-90,width=0.75\textwidth]{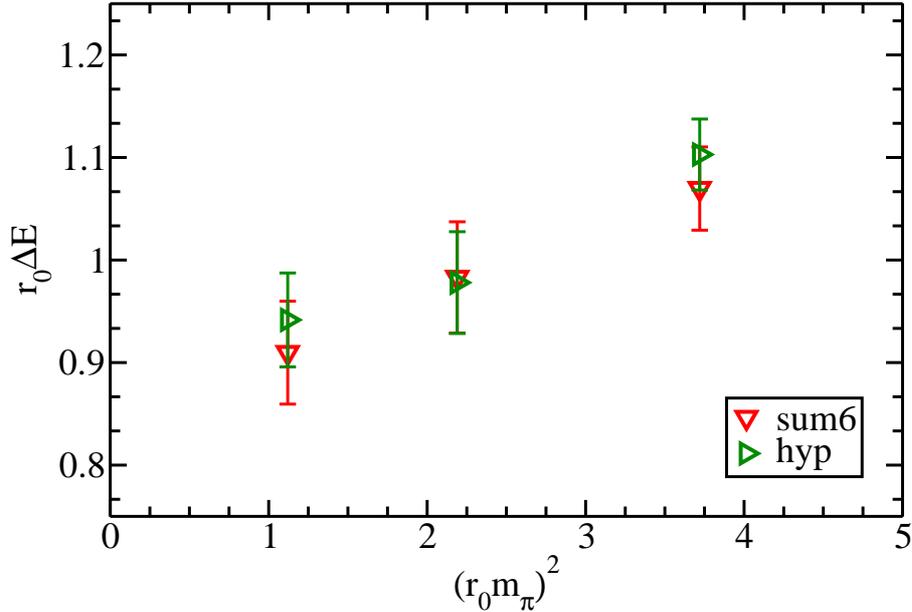}
\caption{(Color online) The energy difference $r_0$[E(1P$_-$)$-$E(1S)] as a function of
the pion mass squared. There is a slight dependence on the light quark mass. Here and
in the following figures for P$_+$, D$_-$ and D$_+$ the results,
from left to right, are from lattices ``DF5'', ``DF4'' and ``DF3'', respectively.}
\label{fig:mpi2_Pm}
\end{figure}

\begin{figure}
\centering
\includegraphics[angle=-90,width=0.75\textwidth]{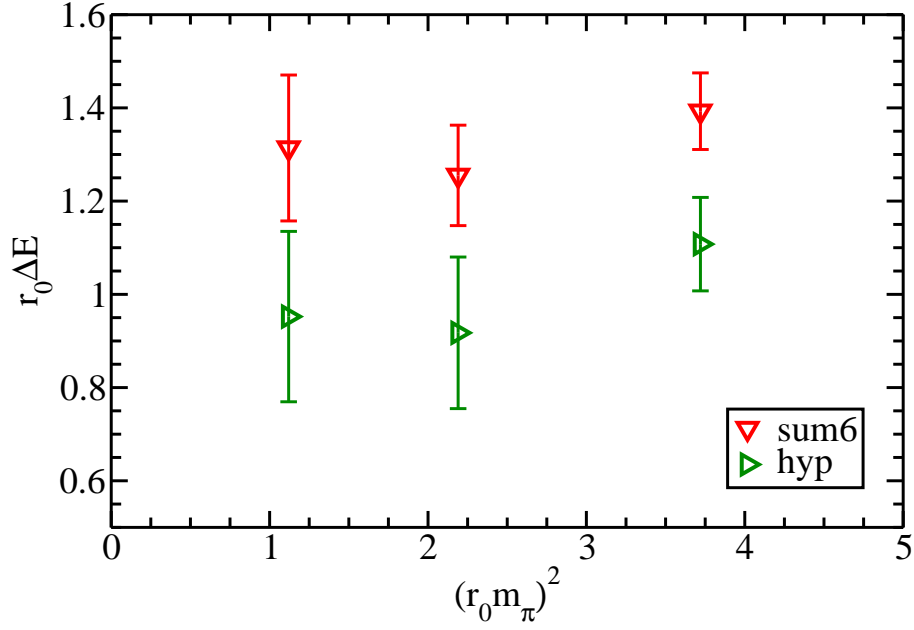}
\caption{(Color online) The energy difference $r_0$[E(1P$_+$)$-$E(1S)] as a function of
the pion mass squared. The dependence on the light quark mass is weak, whereas there
is a manifest difference between ``sum6'' and ``hyp'' configurations.
}
\label{fig:mpi2_Pp}
\end{figure}

\begin{figure}
\centering
\includegraphics[angle=-90,width=0.75\textwidth]{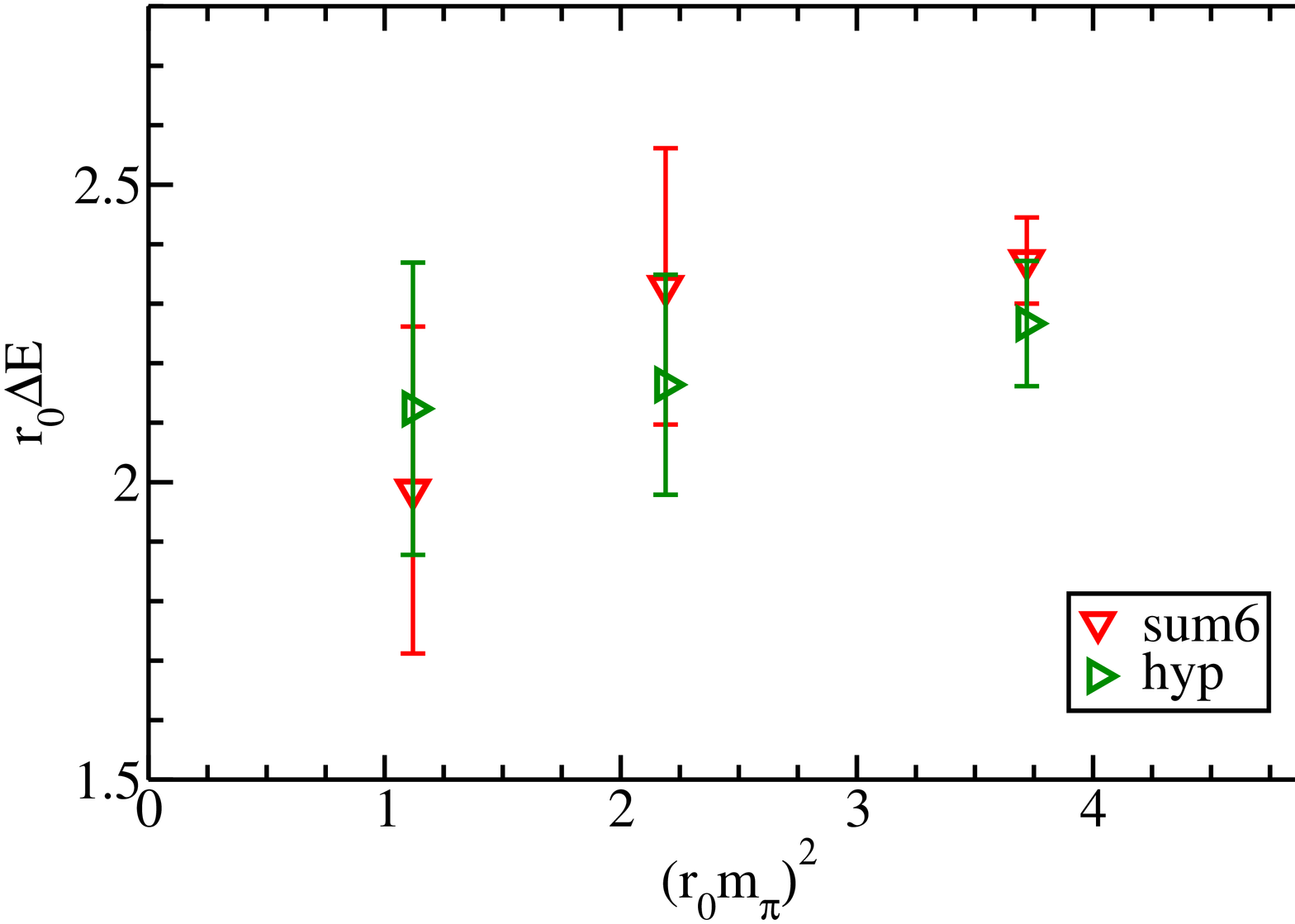}
\caption{(Color online) The energy difference $r_0$[E(1D$_-$)$-$E(1S)] as a function of
the pion mass squared. The dependence on the light quark mass is weak, and the
differences between the ``sum6'' and ``hyp'' smearings are small --- especially
for ``DF3''.}
\label{fig:mpi2_Dm}
\end{figure}

\begin{figure}
\centering
\includegraphics[angle=-90,width=0.75\textwidth]{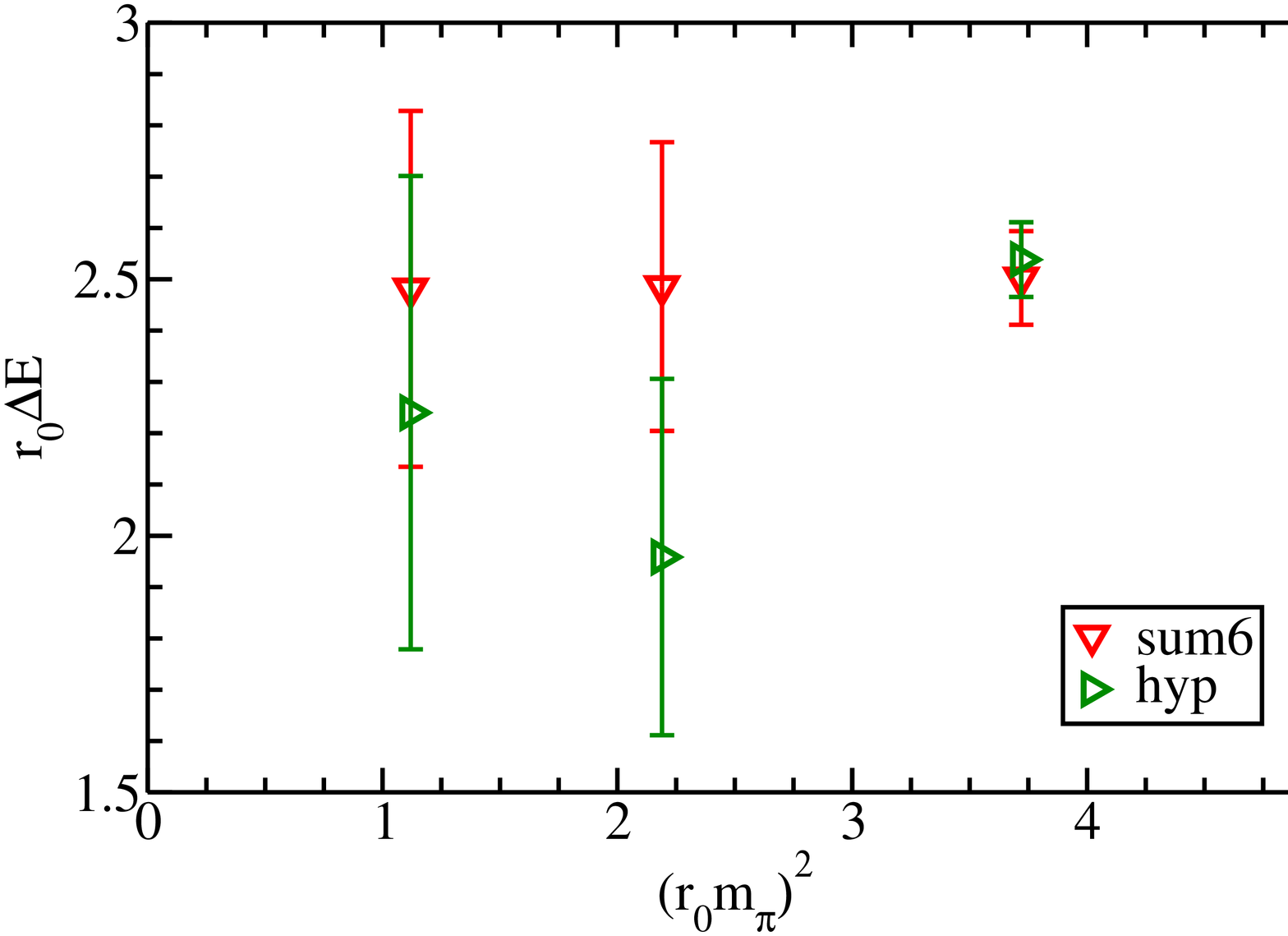}
\caption{(Color online) The energy difference $r_0$[E(1D$_+$)$-$E(1S)] as a function of
the pion mass squared. The dependence on the light quark mass is weak, and the
differences between the ``sum6'' and ``hyp'' smearings are small --- especially
for ``DF3''.}
\label{fig:mpi2_Dp}
\end{figure}

\begin{figure}
\centering
\includegraphics[angle=-90,width=0.95\textwidth]{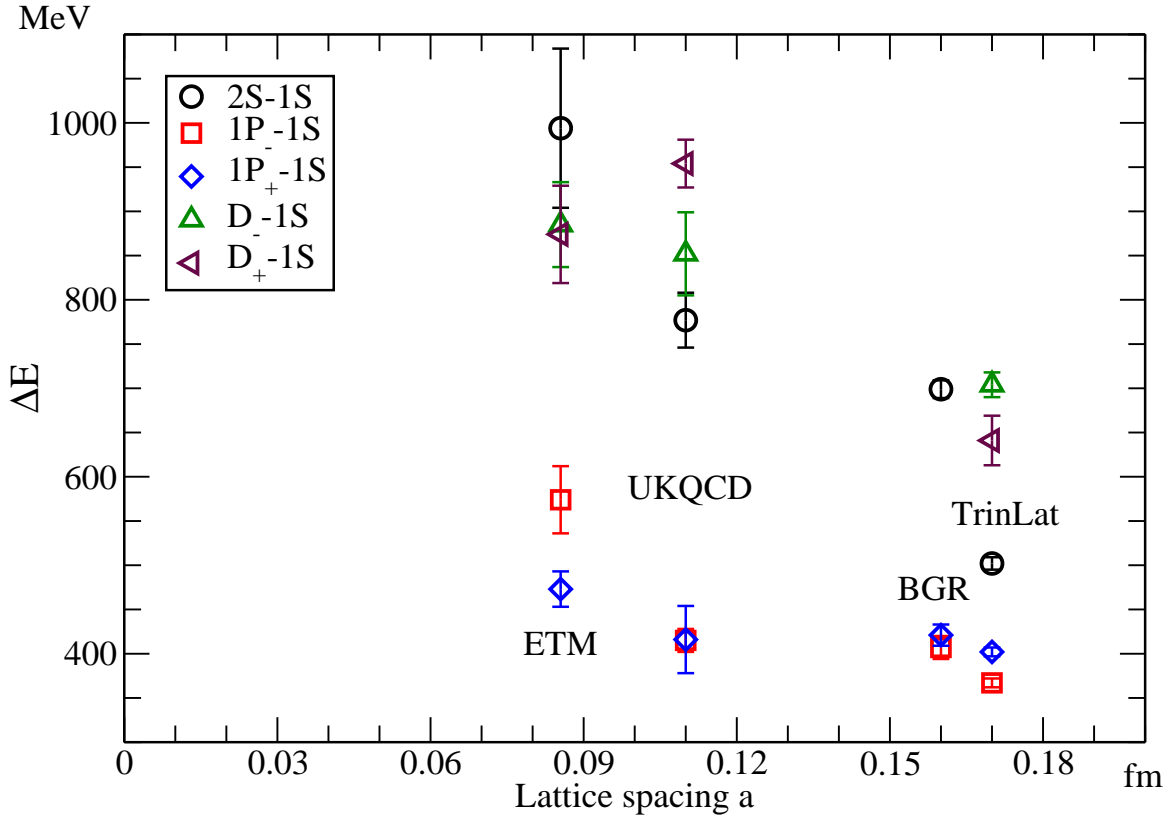}
\caption{(Color online)  Comparison of different static-light lattice results.
On the left at $a=0.0855$~fm are the results from European Twisted Mass
Collaboration \cite{TwistedMass}, at $a=0.11$~fm our results, at
$a=0.16$~fm BGR Collaboration's results \cite{Burch} and on the
right at $a=0.17$~fm TrinLat group's results \cite{II}. A set of similar
lattices with different lattice spacings is needed for going to the continuum
limit.}
\label{fig:latt_comparison}
\end{figure}

To check how the results depend on the light quark mass we plot the
energies (i.e. energy differences with respect to the 1S state) as a
function of the pion mass squared, $(r_0 m_\pi)^2$. As can be seen
in Figs. \ref{fig:mpi2_Pm}--\ref{fig:mpi2_Dp}, for P- and D-wave states
the dependence on the light quark mass is not strong.
We also compare our results to other static-light meson lattice calculations in
Fig.~\ref{fig:latt_comparison}. It is seen that the P-wave results
do not change much between the different lattices, but the 2S-1S and D-wave
energy differences vary a lot. However, since the lattices, quark masses
and lattice spacings are different, the results should only agree in the
continuum limit. A set of similar lattices with different lattice spacings
would be needed to check this.

\subsection{Interpolation to the \textit{b} quark mass}

\begin{figure}
\centering
\includegraphics[angle=-90,width=0.95\textwidth]{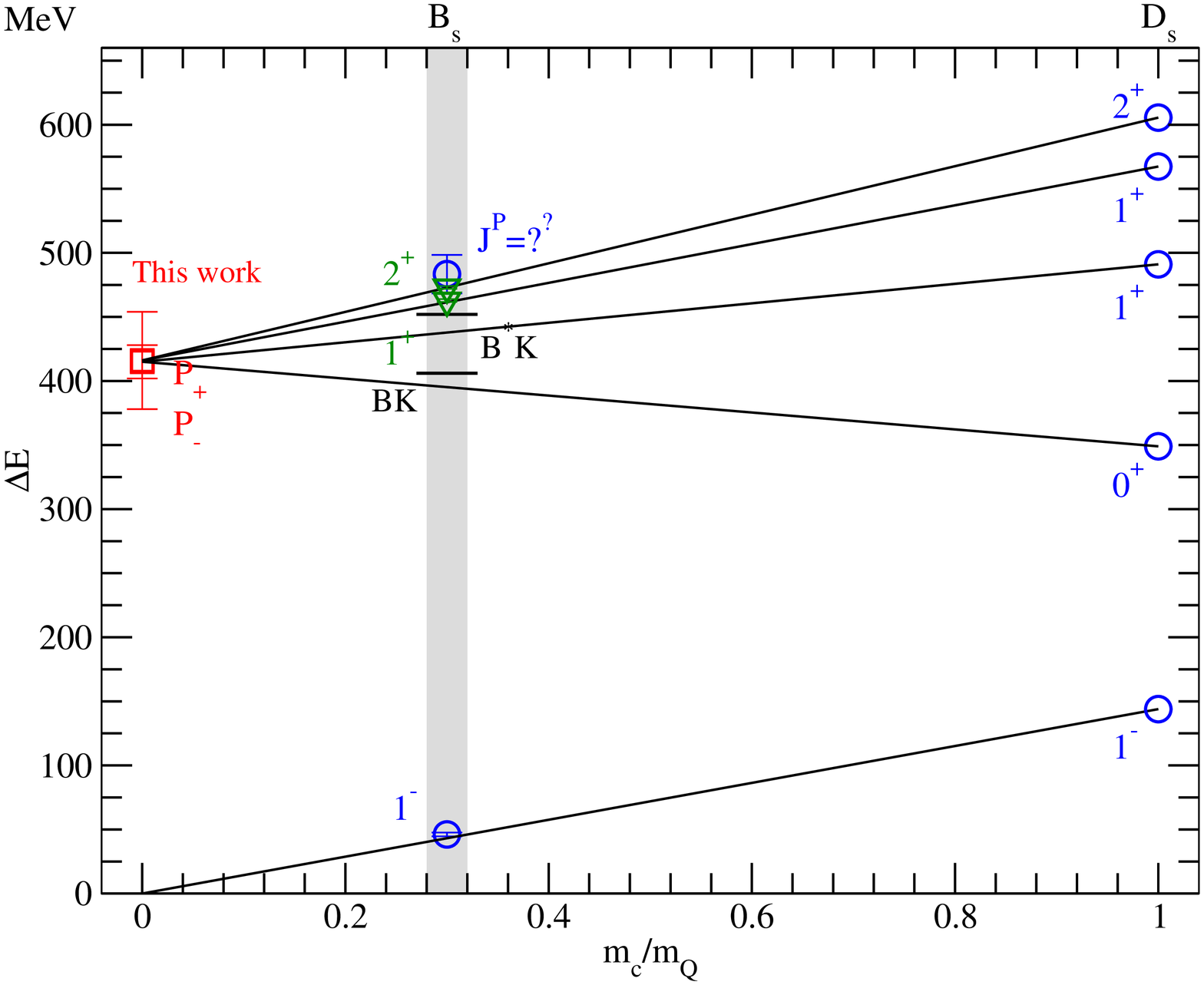}
\caption{(Color online) Interpolation to the \textit{b} quark mass.
The ratio $m_c/m_b$ is taken to be $0.30(2)$ (from~\cite{PDG}; shown by the vertical
band). The $D_s$ meson experimental results are
from \protect\cite{PDG} (blue circles), and the $B_s$ meson experimental results are
from \protect\cite{PDG} (blue circles) and \protect\cite{CDF_D0} (green triangles).
Our results (using ``DF3hyp'' configurations) are marked with red squares.
}
\label{fig:Bs_interpolation}
\end{figure}

Even though we can not go to the continuum limit, we feel that it is
worth-while to try to predict where the $B_s$ meson excited states lie.
To obtain the predictions of the excited state energies, we can now
interpolate in $1/m_Q$, where $m_Q$ is the heavy quark mass, between
the ``DF3'' heavy-light lattice calculations
and $D_s$ meson experimental results, i.e. interpolate
between the static quark ($m_Q=\infty$)  and the charm quark
($m_Q=m_c$). Here we, of course, have to assume that the measured
$D_s$ meson states are simple quark--anti-quark states. This is not
necessarily true: for example the mass of the $D_{s0}^\ast(2317)$ is
much lower than what is predicted by conventional potential models, and it has
thus been proposed that it could be either a four quark state, a $DK$ molecule or
a $D\pi$ atom. A short review on meson excited state spectroscopy and the
puzzles in interpreting the results is given in Ref.~\cite{Poireau}.
However, the inclusion of chiral radiative corrections could change
the potential model predictions considerably~\cite{LeeLee}.
In the following we assume that the states are the usual quark--anti-quark
states.

We use linear interpolation, i.e.
\begin{equation}
\label{linextr}
\Delta E = A + B \frac{m_c}{m_Q} + C F_j \frac{m_c}{m_Q}\, .
\end{equation}
Here $A$, $B$ and $C$ are fit parameters and $F_j =2[J(J+1)-j_q(j_q+1)-s_Q(s_Q+1)]$, where
$J$ is the total angular momentum, $s_Q=1/2$ is the heavy quark spin and $j_q$ is the
combined spin and orbital angular momentum $L$ of the light quark (see Table~\ref{fjTable}).
The interpolation procedure is shown in Fig.~\ref{fig:Bs_interpolation}. Note that the
linear interpolation works perfectly for the $1^-$ S-wave state, where the experimental
energies are known for both $B_s$ and $D_s$ mesons, and the lattice result (zero) is simply
because the two 1S states are automatically degenerate at $m_Q=\infty$.

Our predictions of the energy differences $m(1P)-m(1S)$ for the $B_s$ meson are given in
Table~\ref{BsEneTable}. For our preferred lattice ``DF3hyp'' these agree very well with
the experimental measurements of the energies of the $1^+$ and $2^+$ P-wave states.
There we predict that the two lowest P-wave states lie a few MeV below the $BK$ and $B^\ast K$
thresholds (minus the 1S state energy) at 406 and 452 MeV respectively. We show the ``DFsum6''
results for comparison.

\begin{table}
\centering
 \begin{tabular}{|ccc||ccc||ccc|}
 \hline
 J$^P$ & L & $F_j$ & J$^P$ & L & $F_j$ & J$^P$ & L & $F_j$ \\
 \hline
 $0^+$ & 0 & $-3$ & $0^+$ & 1 & $-3$ & $1^+$ & 1 & $-5$ \\
 $1^+$ & 0 & $+1$ & $1^+$ & 1 & $+1$ & $2^+$ & 1 & $+3$ \\
 \hline
\end{tabular}
\caption{Coefficients $F_j= 2[J(J+1)-j_q(j_q+1)-s_Q(s_Q+1)]$ (equation~\ref{linextr}).}
\label{fjTable}
\end{table}

\begin{table}
\centering
 \begin{tabular}{|c||c|c|c|}
 \hline
 J$^P$ & DF3hyp & DF3sum6   & experiments \\
 \hline
 $0^+$ & $393 \pm  9$ MeV & $384 \pm 10$ MeV & -  \\
 $1^+$ & $440 \pm  9$ MeV & $432 \pm 10$ MeV & -  \\
 \hline			                     
 $1^+$ & $466 \pm 25$ MeV & $538 \pm 21$ MeV & $463 \pm 1$ MeV \\
 $2^+$ & $482 \pm 25$ MeV & $551 \pm 21$ MeV & $473 \pm 1$ MeV \\
 \hline
\end{tabular}
\caption{Our predictions for $B_s$ meson mass differences, M($B^\ast_s$)-M($B_s$),
for the P-wave states. The uncertainty in the ratio $m_c/m_b$ is not taken into
account in the error estimates. The experimental results are from \cite{CDF_D0}.
}
\label{BsEneTable}
\end{table}

\begin{figure}
\centering
\includegraphics[angle=-90,width=0.95\textwidth]{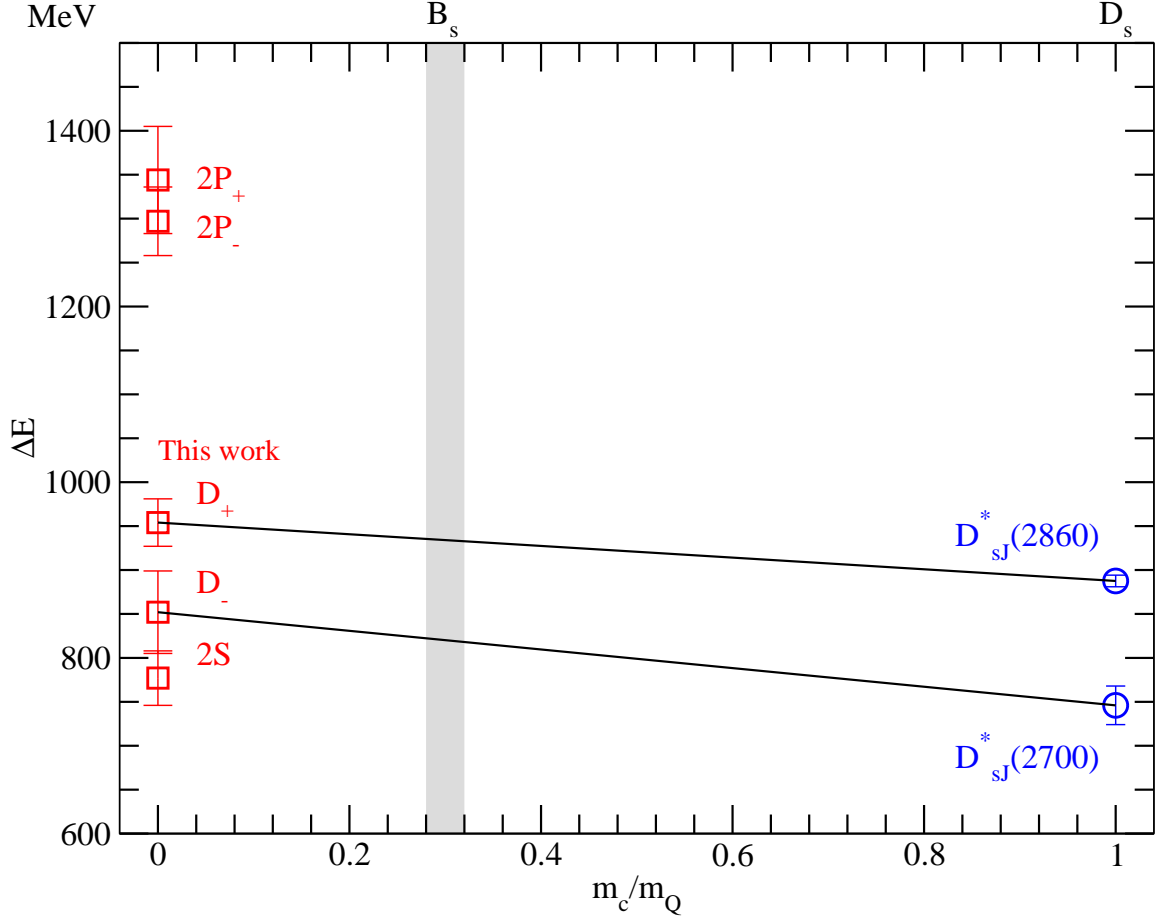}
\caption{(Color online) Interpolation to the \textit{b} quark mass for
``DF3hyp'' lattice: higher excited states. The lines illustrate what
the interpolation would look like, if the $D_s$ meson states were
D-wave states. The experimental results are
from~\protect\cite{Aubert:2006mh, Abe:2006xm}.
Interpolating to $m_c/m_b$ predicts D-wave $J^P=1^{-},~3^{-}$
at 817(31) and 932(18) MeV respectively.}
\label{fig:Bs_interp2}
\end{figure}

As for other excited states, BaBar and Belle observed two new states,
$D_{sJ}^\ast(2860)$ and $D_{sJ}^\ast(2700)$, in 2006
\cite{Aubert:2006mh, Abe:2006xm}. The $J^P$ quantum numbers of the
$D_{sJ}^\ast(2860)$ can be $0^+$, $1^-$, $2^+$, etc., so it could
be a radial excitation of the $D_{s0}^\ast(2317)$ or a $J^P=3^-$
D-wave state. The first interpretation is rather popular, but our lattice
results favor the D-wave $J^P=3^-$ assignment in  agreement with
Colangelo, De Fazio and Nicotri \cite{Colangelo}.
(The slope of the interpolating line would be very steep, if the $D_{sJ}^\ast(2860)$
is a radial excitation of the $D_{s0}^\ast(2317)$.)
Interpolation then predicts a D-wave $J^P=3^-$ $B_s$ state at 932(18) MeV.
In addition, the $D_{sJ}^\ast(2700)$ could be a radially excited S-wave state or a
D-wave $J^P=1^-$ state. If the latter identification is assumed,
then a D-wave $J^P=1^-$ $B_s $ state at 817(31) MeV is expected (see
Fig.~\ref{fig:Bs_interp2}).

\subsection{Bayesian ideas}
\label{bayes}

\begin{figure}
\centering
\includegraphics[height=0.65\textwidth, angle=-90]{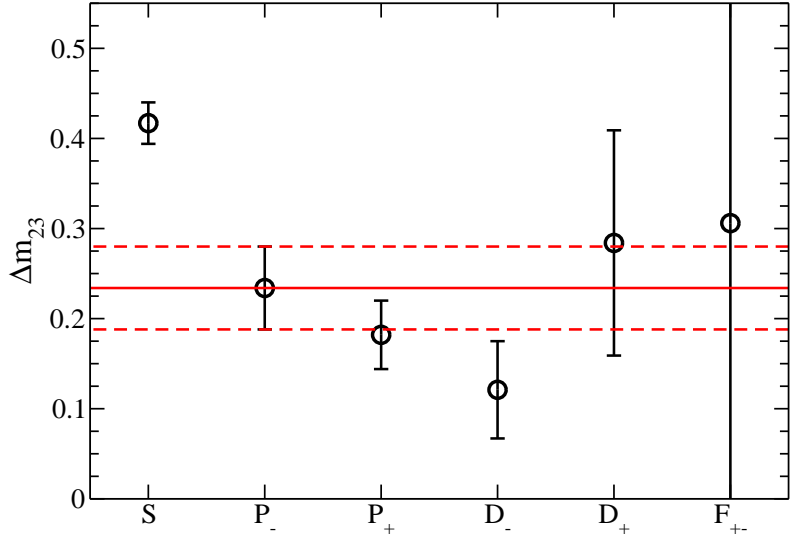}
\caption{(Color online) Mass differences $\Delta m_{32}=m_3-m_2$ for the ``DF3hyp''
data in lattice units. This seems to be almost constant for angular
momentum L$\ge 1$ (given the sizeable errors on the data). The lines give
the $\Delta m_{32}=0.234(46)$ that is used in this study. Looking at the ``DF3sum6''
data gives a very similar picture (not shown here) and an estimate
of $\Delta m_{32}=0.207(54)$. See section \ref{bayes} for details.}
\label{fig:bayes}
\end{figure}

\begin{table}
\centering
 \begin{tabular}{|c|ccccc|}
 \hline
 nL$_{\pm}$ &\, $m_3$ 3 exp\, &\, $m_{3,\textrm{ prior}}$\, &\, $m_3$ Bayes\, &\,  $m_2$ 3 exp\, &\, $m_2$ Bayes\, \\
 \hline
 P$_+$    & 1.46(2)  & 1.52(6) & 1.471(15)& 1.28(3)   & 1.29(3) \\
 D$_-$    & 1.63(5)  & 1.74(6) & 1.67(4)  & 1.51(3)   & 1.52(2) \\
 D$_+$    & 1.84(12) & 1.79(5) & 1.80(2)  & 1.559(11) & 1.558(8)\\
 F$_{+-}$ & 1.96(43) & 1.89(5) & 1.89(1)  & 1.66(2)   & 1.657(5)\\
 \hline
\end{tabular}
\caption{
Comparison of the $m_{3,\textrm{prior}}$ with the results from the full 3 exponential fit and
the Bayesian fit for ``DF3hyp'' configurations.
See Section~\ref{bayes} for definition of $m_{3,\textrm{prior}}$. ``$m_3$ 3 exp'' and
``$m_3$ Bayes'' are the results of a full 3 exponential fit and a Bayesian (fixed $m_{32}$)
fit, respectively. Likewise for the $m_2$.
The P$_+$ ``Bayes'' fit is merely to check
that the Bayesian ideas work well and does not restrict the analysis too much.}
\label{BayesTable_hyp}
\end{table}

\begin{table}
\centering
 \begin{tabular}{|c|ccccc|}
 \hline
 nL$_{\pm}$ &\, $m_3$ 3 exp\, &\, $m_{3,\textrm{ prior}}$\, &\, $m_3$ Bayes\, &\,  $m_2$ 3 exp\, &\, $m_2$ Bayes\, \\
 \hline
 P$_+$    & 1.59(2) & 1.61(5) & 1.59(2)  & 1.39(2)   & 1.393(15) \\
 D$_-$    & 1.80(9) & 1.79(5) & 1.79(3)  & 1.578(11) & 1.577(8)  \\
 D$_+$    & 2.1(2)  & 1.82(5) & 1.84(2)  & 1.604(12) & 1.596(11) \\
 F$_{+-}$ & 2.6(9)  & 1.94(6) & 1.950(11)& 1.72(2)   & 1.721(8)  \\
 \hline
\end{tabular}
\caption{Comparison of the $m_{3,\textrm{prior}}$ with the results from the full 3 exponential fit and
the Bayesian fit for ``DF3sum6'' configurations. Again, the prior $m_3$ values are in fairly
good agreement with the $m_3$ results from the full 3 exponential fits. Fixing $m_3-m_2$
does not change the first excited state $m_2$. See Table~\protect\ref{BayesTable_hyp} for
notation.}
\label{BayesTable_sum6}
\end{table}

In some cases, using 3 exponentials to fit the $C_2$ data does not work very well.
In Table~\ref{C2FitTable}, these cases are marked with an asterisk.
Since these fits are not as good or stable as one would hope, we introduce some
Bayesian ideas and use prior knowledge of the energies to constrain the fit,
or rather to guide the fit in the right direction.
The third mass, $m_3$, (which would be the mass of the second radial excitation,
if there was no pollution from higher states) is restricted to be in the range
$m_{3,\textrm{ prior}}\pm \Delta m_{3,\textrm{ prior}}$ by adding a term
\begin{equation}
\frac{(m_3-m_{3,\textrm{ prior}})^2}{(\Delta m_{3,\textrm{ prior}})^2}
\label{eq:m3prior}
\end{equation}
to the $\chi^2$. This is not a hard constraint, unlike fixing $m_3$ to a given value would be,
but rather constrains the parameter to a given range softly. The $m_{3,\textrm{ prior}}$ and
$\Delta m_{3,\textrm{ prior}}$ are determined beforehand by estimating the difference
$\Delta m_{32}=m_3-m_2$ from full 3 exponentials fits. This mass difference seems to be almost
constant for states that have $L=1$ or higher (see Fig.~\ref{fig:bayes}). Therefore we use the
P$_-$ state to set the $\Delta m_{32}$ for D-wave and F-wave states. The $m_{3,\textrm{ prior}}$
for D-wave and F-wave states is then calculated by adding $\Delta m_{32}$ to the $m_2$ from
the full 3 exponential fit for the state in question
(see Tables~\ref{BayesTable_hyp},~\ref{BayesTable_sum6}).
The prior $m_3$ values are in fairly good agreement with the $m_3$ results from the full
3 exponential fits, and fixing $m_3-m_2$ does not change the first excited state $m_2$.
The P$_+$ ``Bayes'' fit are used to check that the Bayesian ideas work well and does not
restrict the analysis too much.

\subsection{Spin-orbit splitting}

\begin{figure}
\centering
\includegraphics[height=0.65\textwidth, angle=-90]{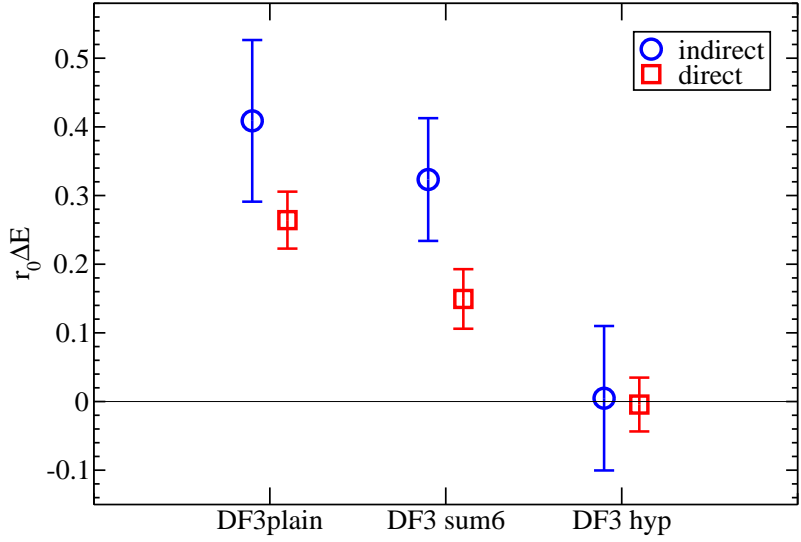}
\caption{(Color online) The Spin-Orbit splittings of P-wave states for the
``DF3'' lattice.
}
\label{fig:SOSPDF3}
\end{figure}

\begin{figure}
\centering
\includegraphics[height=0.65\textwidth, angle=-90]{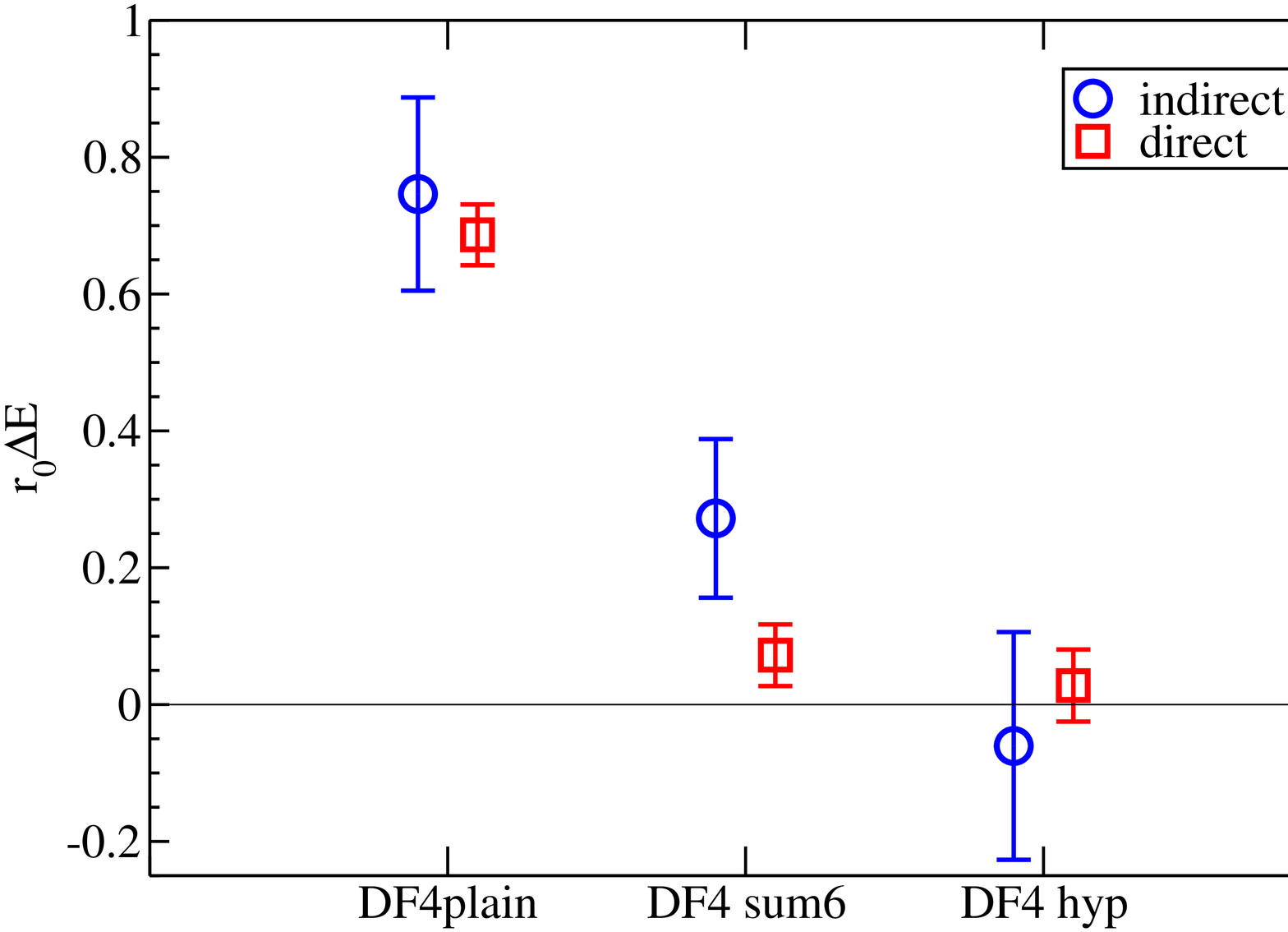}
\caption{(Color online) The Spin-Orbit splittings of P-wave states for the
``DF4'' lattice.
}
\label{fig:SOSPDF4}
\end{figure}

\begin{figure}
\centering
\includegraphics[height=0.65\textwidth, angle=-90]{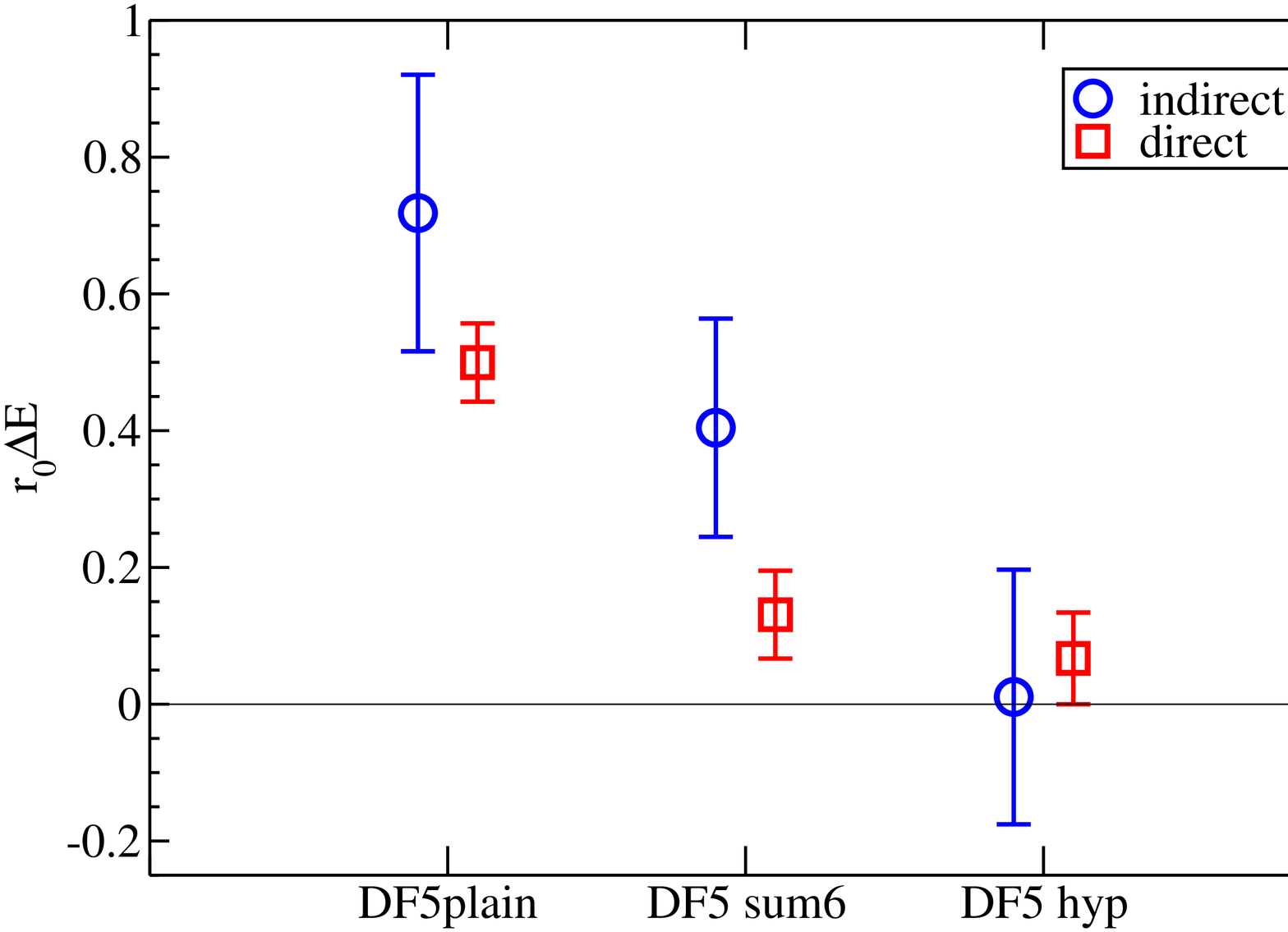}
\caption{(Color online) The Spin-Orbit splittings of P-wave states for the
``DF5'' lattice.
}
\label{fig:SOSPDF5}
\end{figure}

\begin{figure}
\centering
\includegraphics[height=0.65\textwidth, angle=-90]{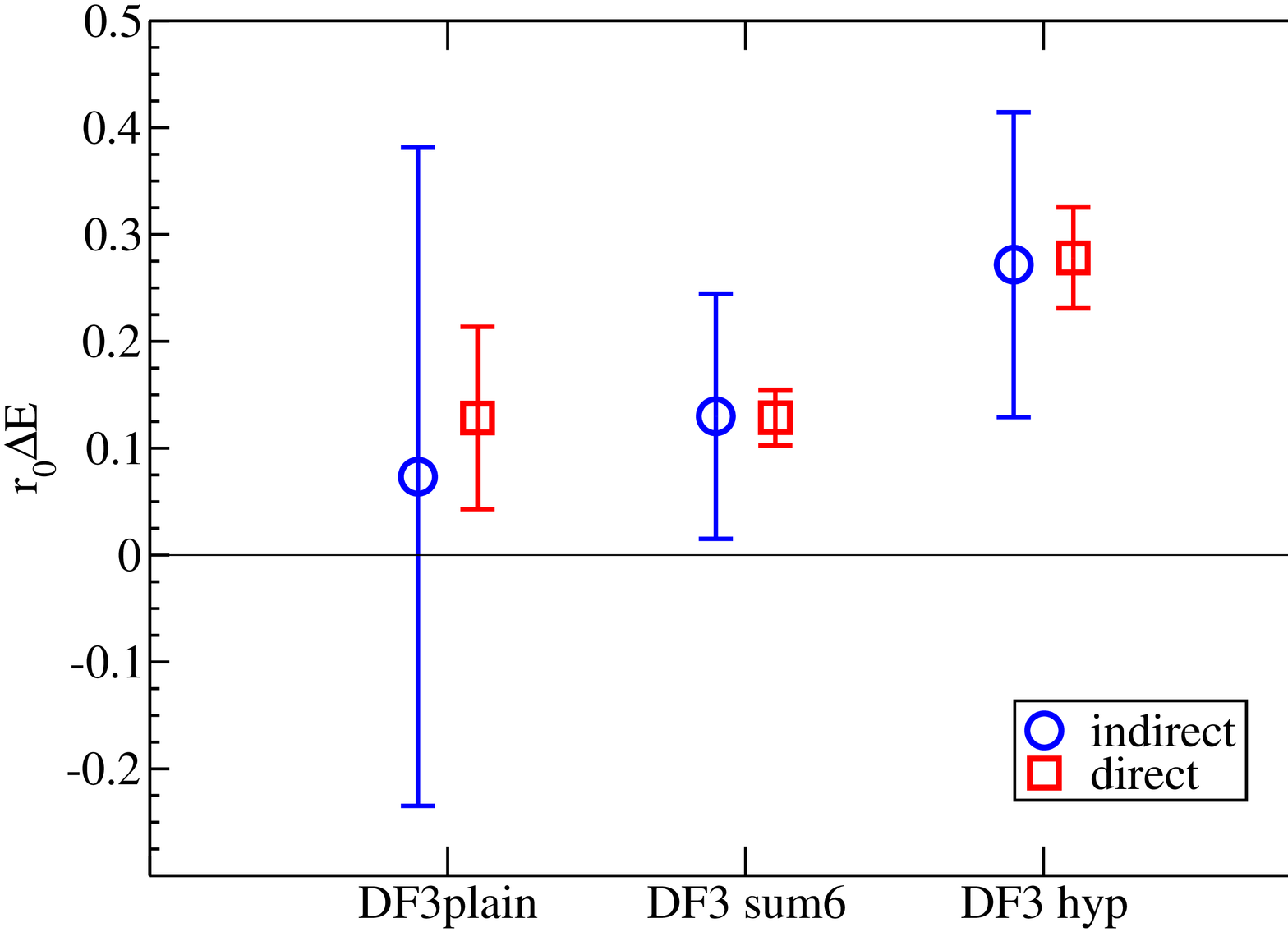}
\caption{(Color online) The Spin-Orbit splittings of D-wave states for the
``DF3'' lattice.
}
\label{fig:SOSDDF3}
\end{figure}

\begin{figure}
\centering
\includegraphics[height=0.65\textwidth, angle=-90]{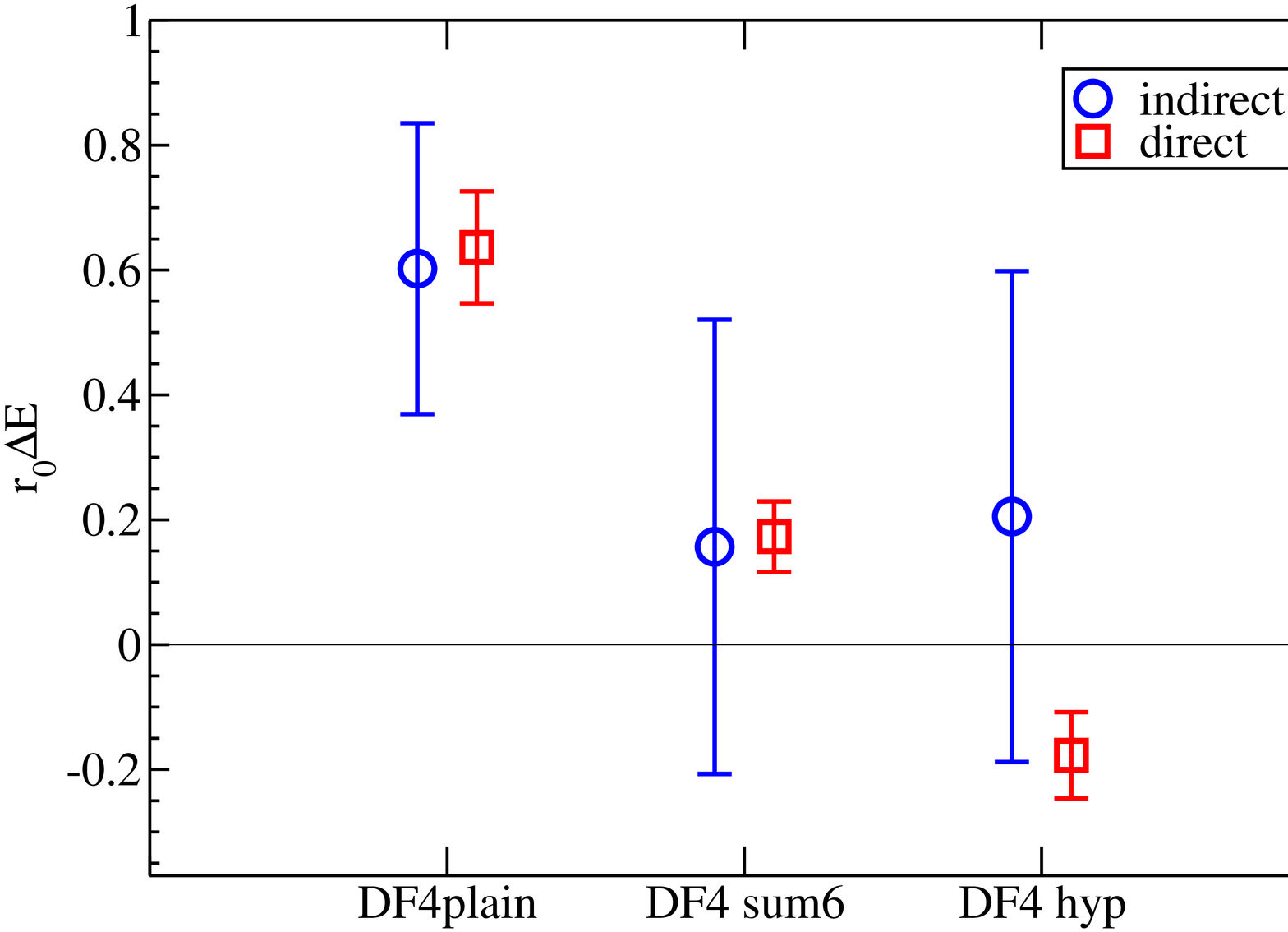}
\caption{(Color online) The Spin-Orbit splittings of D-wave states for the
``DF4'' lattice.
}
\label{fig:SOSDDF4}
\end{figure}

\begin{figure}
\centering
\includegraphics[height=0.65\textwidth, angle=-90]{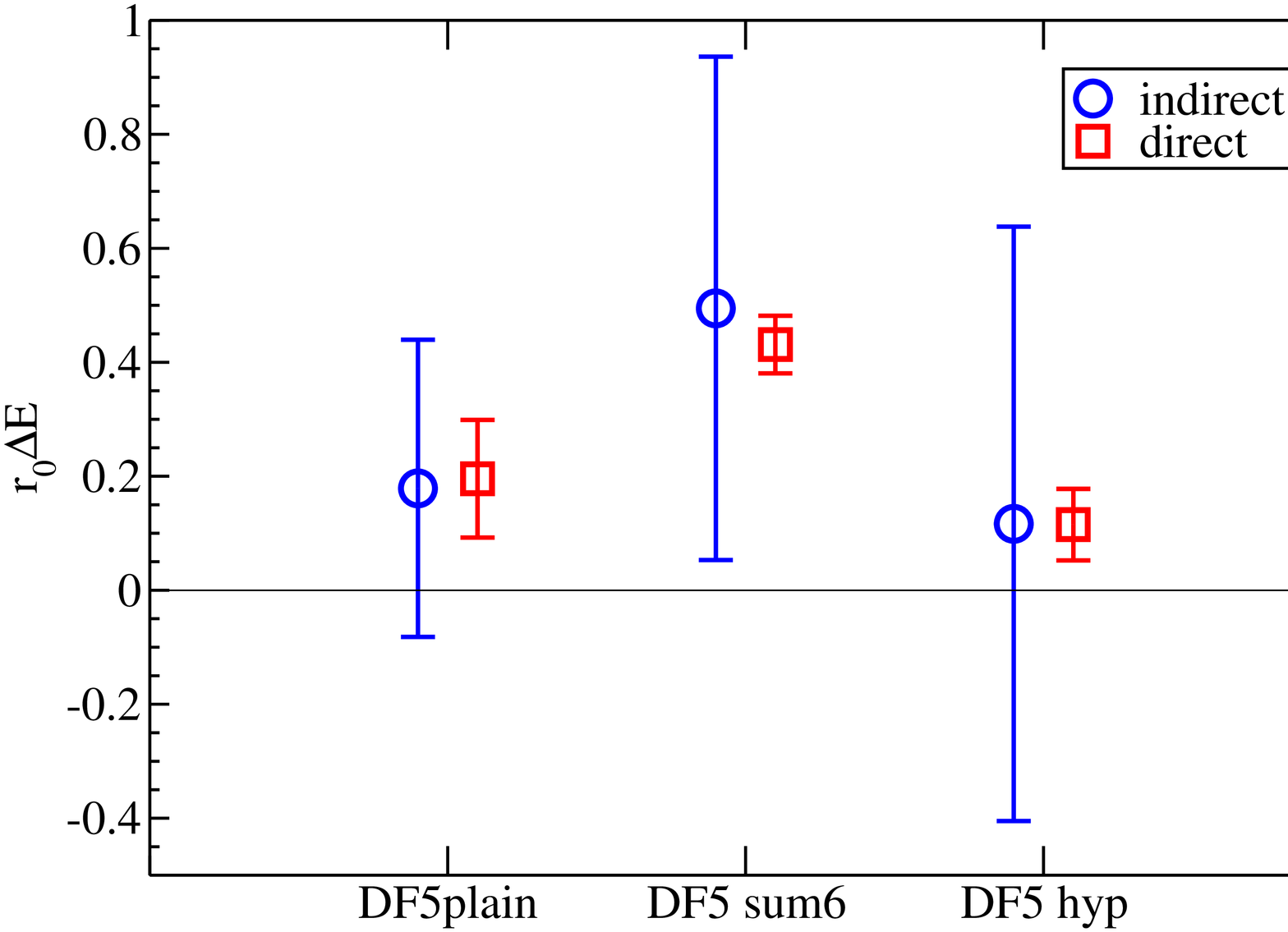}
\caption{(Color online) The Spin-Orbit splittings of D-wave states for the
``DF5'' lattice.
}
\label{fig:SOSDDF5}
\end{figure}

\begin{table}
\centering
 \begin{tabular}{|c|c|c||c|c|c||c|c|c|}
 \hline
 Lattice & Direct  & Indirect & Lattice  & Direct  & Indirect  & Lattice  & Direct  & Indirect \\
 \hline
DF3plain & 0.26(4) & 0.41(12) & DF4plain & 0.69(4) &  0.75(14) & DF5plain & 0.50(6) & 0.7(2)   \\
DF3sum6  & 0.15(4) & 0.32(9)  & DF4sum6  & 0.07(5) &  0.27(12) & DF5sum6  & 0.13(6) & 0.40(16) \\
DF3hyp   & 0.00(4) & 0.00(11) & DF4hyp   & 0.03(5) & -0.06(17) & DF5hyp   & 0.07(7) & 0.0(2)   \\
 \hline
\end{tabular}
\caption{P-wave spin-orbit splitting $r_0\Delta E=r_0 (m_\textrm{1P$_+$}-m_\textrm{1P$_-$})$
for the different lattices. To get $\Delta E$ in GeV requires a factor 0.38(2).}
\label{SOSTableP}
\end{table}

\begin{table}
\centering
 \begin{tabular}{|c|c|c||c|c|c||c|c|c|}
 \hline
 Lattice & Direct  & Indirect & Lattice  & Direct   & Indirect & Lattice  & Direct  & Indirect \\
 \hline
DF3plain & 0.13(9) & 0.1(3)   & DF4plain &  0.64(9) & 0.6(2)   & DF5plain & 0.20(10) & 0.2(3)   \\
DF3sum6  & 0.13(3) & 0.13(11) & DF4sum6  &  0.17(6) & 0.2(4)   & DF5sum6  & 0.43(5)  & 0.5(4)   \\
DF3hyp   & 0.28(5) & 0.27(14) & DF4hyp   & -0.18(7) & 0.2(4)   & DF5hyp   & 0.12(6)  & 0.1(5)   \\
 \hline
\end{tabular}
\caption{D-wave spin-orbit splitting $r_0\Delta E=r_0 (m_\textrm{1D$_+$}-m_\textrm{1D$_-$})$
for the different lattices. To get $\Delta E$ in GeV requires a factor 0.38(2).}
\label{SOSTableD}
\end{table}

One interesting point to note here is that the spin-orbit splitting of the P-wave states
is small, almost zero, for the preferred ``hyp'' smearing. We extract this energy difference
of the 1P$_+$ and 1P$_-$ states in two different ways:
\begin{enumerate}
\item
\emph{Indirectly} by simply calculating the difference using the energies given by the fits
in Eq.~\ref{C2fit}, when the P$_+$ and P$_-$ data are fitted separately.

\item
Combining the P$_+$ and P$_-$ data and fitting the ratio
$\textrm{C}_2(\textrm{P}_+)/\textrm{C}_2(\textrm{P}_-)$, which enables us
to go \emph{directly} for the spin-orbit splitting, $m_\textrm{1P$_+$}-m_\textrm{1P$_-$}$.
\end{enumerate}
In the latter case, the expression (for a given fuzzing) is
\begin{equation}
\frac{\textrm{C}_2(\textrm{P}_+)}{\textrm{C}_2(\textrm{P}_-)}=
A\, \mathrm{e}^{-\Delta m_1\, T}\bigg[
\frac{1+b_2^+\, \mathrm{e}^{-\Delta m_2^+\, T}+b_3^+\, \mathrm{e}^{-\Delta m_3^+\, T}}{
1+b_2^-\, \mathrm{e}^{-\Delta m_2^-\, T}+b_3^-\, 
\mathrm{e}^{-\Delta m_3^-\, T}}\bigg],
\end{equation}
where
\begin{align*}
\Delta m_1  =&\; m_\textrm{1P$_+$}-m_\textrm{1P$_-$}\, ,\\
\Delta m_2^+=&\; m_\textrm{2P$_+$}-m_\textrm{1P$_+$}\, ,\\
\Delta m_2^-=&\; m_\textrm{2P$_-$}-m_\textrm{1P$_-$}\, ,\\
\Delta m_3^+=&\; m_\textrm{3P$_+$}-m_\textrm{1P$_+$}\textrm{ and}\\
\Delta m_3^-=&\; m_\textrm{3P$_-$}-m_\textrm{1P$_-$}.
\end{align*}
We get the best results
by fitting $\Delta m_1$, $\Delta m_2^+$, $\Delta m_2^-$ and the coefficients $A$,
$b_2^+$ and $b_2^-$, but fixing the remaining mass differences and $b$'s from the
individual two-point correlator fits (equation~\ref{C2fit} and Table~\ref{C2FitTable}). Thus
\begin{equation}
b_3^+ = \frac{a_{3,f_1}(\textrm{P}_+)a_{3,f_2}(\textrm{P}_+)}{a_{1,f_1}(\textrm{P}_+)a_{1,f_2}(\textrm{P}_+)}
\textrm{ and }
b_3^- = \frac{a_{3,f_1}(\textrm{P}_-)a_{3,f_2}(\textrm{P}_-)}{a_{1,f_1}(\textrm{P}_-)a_{1,f_2}(\textrm{P}_-)}
\end{equation}
for given values of fuzzing indices $f_1$, $f_2$.
The D-wave spin-orbit splitting is also extracted in a similar manner.
The results of the fits are given in Tables~\ref{SOSTableP},~\ref{SOSTableD}
and in Figs.~\ref{fig:SOSPDF3}--\ref{fig:SOSDDF5}. In all cases the errors on
the direct estimates are much smaller than those on the indirect ones.
Also in most cases the direct and indirect estimates are consistent with each
other --- the only exception being the P-wave ``sum6'' estimates. There the
direct value is somewhat lower than the indirect estimate. In fact this
difference brings the ``sum6'' direct estimate closer to the ``hyp'' value,
and lends support to the preferred ``hyp'' estimate, which in all three cases
gives a small P-wave spin-orbit splitting (SOS), consistent with zero, for
the ``hyp'' configurations. The D-wave spin-orbit splitting (SOS) results are
more varied, but the ``DF3hyp'' lattice suggests clearly a positive, non-zero
D-wave SOS. However, the ``DF4hyp'' and ``DF5hyp'' estimates are considerably
smaller, becoming negative for the ``DF4hyp''. At present it is not clear
whether this is a lattice artefact due to, say, not being in the continuum
limit, or that indeed the D-wave results are more dependent on $m_q$ than
in the P-wave case.

\section{A model based on the Dirac equation}
\label{diracmodel}

Since the mass of the heavy quark is infinite, we have for a potential
description essentially a one-body problem. Therefore, a simple model based
on the Dirac equation is used to try to describe the lattice data. The
potential in the Dirac equation has the usual  linearly rising scalar
part, $b_{\textrm{sc}} r$, but in addition a vector part $b_{\textrm{vec}} r$
is added. The one gluon exchange term, $a_{\textrm{OGE}}\cdot V_{\textrm{OGE}}$,
where
\begin{equation}
V_{\textrm{OGE}}=-\frac{4}{3}\frac{\alpha_s(r)}{r},
\end{equation}
with the running coupling constant $\alpha_s(r)$ given by
\begin{equation}
\alpha_s(r)=\frac{2}{\pi} \int_0^{\infty}\!\! \mathrm{d}k\,
\frac{\sin (kr)}{k}\alpha_s(k^2)
\end{equation}
and
\begin{equation}
\alpha_s(k^2)=\frac{12 \pi}{27}\frac{1}{\ln [(k^2+4 m^2_g)/(\Lambda^2_{\textrm{QCD}})]}.
\end{equation}
Here, guided by fits to various meson masses using the Blankenbecler--Sugar
equation, we fix   $\Lambda_{\textrm{QCD}}=260$~MeV and
the dynamical gluon mass $m_g=290$~MeV (see~\cite{lahde} for details).
The potential also has a scalar term $m\omega L(L+1)$, which 
seems to be  needed to increase the energy of D-wave states. This type of
term  arises in flux tube models, where a flux tube's
rotational energy is proportional to $L(L+1)$ (like in Isgur--Paton flux
tube model, \cite{fluxtube}).

\begin{figure}
\centering
\includegraphics[height=0.70\textwidth, angle=-90]{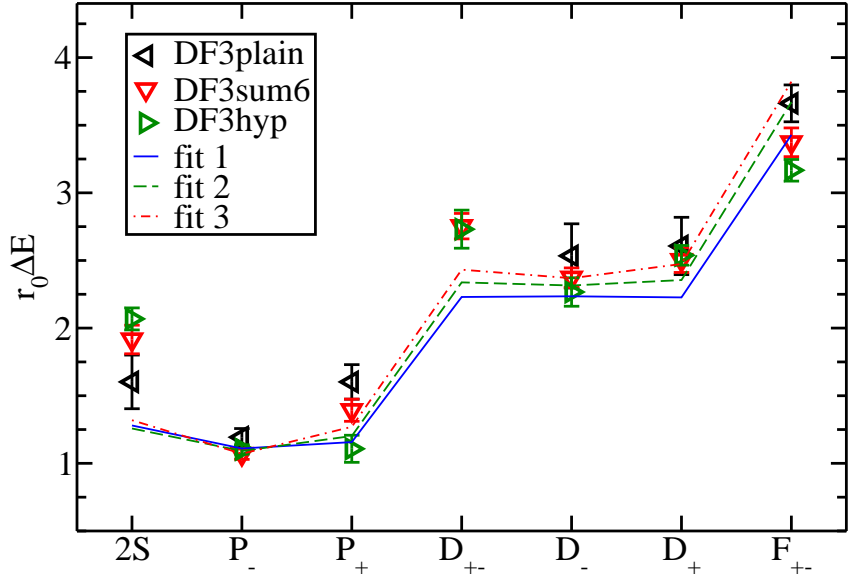}
\caption{(Color online) Energy spectrum of the heavy-light meson and three Dirac model
fits. ``Fit 1'' is a fit to ``DF3hyp'' 1P$_-$, SOS(1P) and 1D$_-$, whereas ``fit 2'' is an
attempt to fit ``DF3hyp'' 1P$_-$, SOS(1P), 1D$_-$ and SOS(1D) (see Table~\ref{DiracTable}).
``Fit 3'' is a fit to ``DF3sum6'' 1P$_-$, SOS(1P), 1D$_-$ and SOS(1D), and is shown here for
comparison.}
\label{fig:edirac}
\end{figure}

\begin{figure}
\centering
\includegraphics[height=0.70\textwidth, angle=-90]{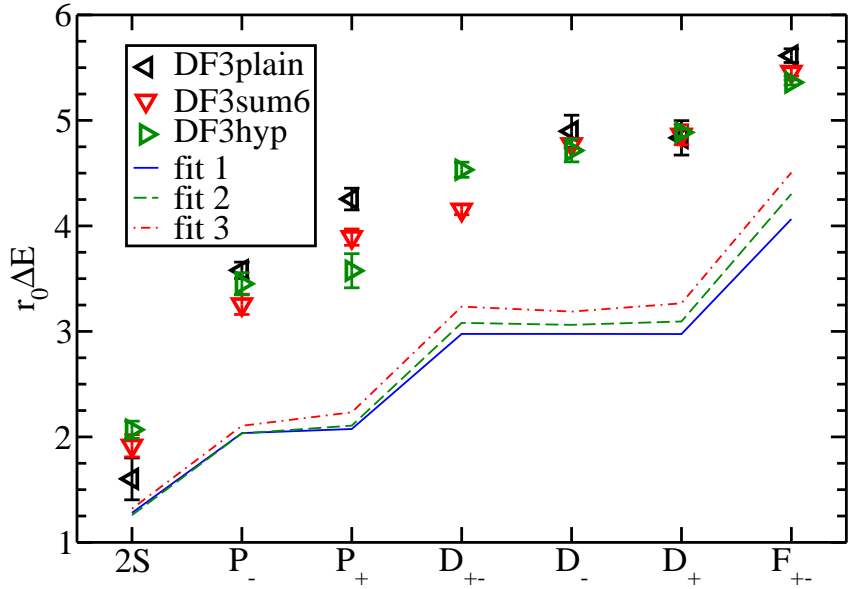}
\caption{(Color online) Energies of the first radial excitations of the heavy-light meson
and the same Dirac model fits shown in Fig.~\ref{fig:edirac}.}
\label{fig:edirac2}
\end{figure}

The lines in the energy spectrum plot (Fig.~\ref{fig:edirac}) show three
Dirac model fits from Table~\ref{DiracTable} with $m = 560$~MeV
(the constituent quark mass, from~\cite{lahde}) and
$a_{\textrm{OGE}} = 1.00$. Attempts to also vary $a_{\textrm{OGE}}$ easily
lead to instabilities. The solid line, labelled ``fit 1'', is a fit
to three ``DF3hyp'' energy differences: 1P$_-$ and 1D$_-$ with respect to the
ground state, and the P-wave spin-orbit splitting (direct estimate) SOS(1P)
[i.e. E(1P$_+$)$-$E(1P$_-$)]. The fit to these energies is acceptable with
total $\chi^2=1.68$, but as soon as a fourth state [e.g. SOS(D)] is added
a good $\chi^2$ can no longer be achieved. The dashed line, ``fit 2'', shows
an attempt to fit ``DF3hyp'' 1P$_-$, SOS(1P), 1D$_-$ and SOS(1D). The $\chi^2$
is not good, and letting $a_{\textrm{OGE}}$ vary does not help: that only
leads to unphysical values for the parameters. Using a different constituent
quark mass, say $m = 490$~MeV from~\cite{Schnitzer}, gives basically the same
fits (the changes are minimal). ``Fit 3'' is a fit to ``DF3sum6'' 1P$_-$,
SOS(1P), 1D$_-$ and SOS(1D), and is shown in the figure for comparison.
The fits to ``DF3sum6'' energies are also shown in Table~\ref{DiracTable}.
In Fig.~\ref{fig:edirac2} the same Dirac model fits are shown for the excited
states. Here it can be seen that the fit is about 500~MeV lower than the
lattice results, and the shift seems to be constant for both lattices (``DF3sum6''
and ``DF3hyp'') for all states, except the 2S. There is no obvious reason why
the Dirac model should underestimate the first radial excitations by a constant
amount, but a term of the form $0.5 (n-1)$~GeV could be included in the model
to improve the fit to excited states and  be interpreted as a flux tube effect
in the same philosophy as the $\omega L(L+1)$ term. However, as the fit to
the ground state energies is poor, this improvement is not pursued.

\begin{table}
\centering
 \begin{tabular}{|l|cccc|}
 \hline
Case &  $b_{\textrm{sc}}$ [GeV/fm] & $b_{\textrm{vec}}$  [GeV/fm] & $\omega$ & total $\chi^2$\\
 \hline
sum6 P$_-$, SOS(1P)                & 1.168(2)   & 0.0       & 0.0        & 7.8 \\
sum6 P$_-$, SOS(1P)                & 0.86(13)   & 0.57(18)  & 0.0        & 0.0 \\
sum6 P$_-$, SOS(1P), D$_-$         & 0.372(14)  & 0.0       & 0.0916(14) & 4.5 \\
sum6 P$_-$, SOS(1P), D$_-$, SOS(1D)& 0.265(4)  & 0.571(13)  & 0.0696(11) & 2.1 \\
 \hline
hyp P$_-$, SOS(1P)                 & 1.294(10)  &  0.0      & 0.0        & 0.54 \\
hyp P$_-$, SOS(1P)                 & 1.40(18)   & -0.2(2)   & 0.0        & 0.00 \\
hyp P$_-$, SOS(1P), D$_-$          & 0.763(9)   &  0.0      & 0.0554(9)  & 1.83 \\
hyp P$_-$, SOS(1P), D$_-$, SOS(1D) & 0.48(5)    &  0.26(2)  & 0.066(5)   & 34.3 \\
 \hline
\end{tabular}
\caption{Dirac model fits for ``DF3''. Here $a_{\textrm{OGE}}=1$ and constituent
quark mass $m = 560$~MeV.
Fits are attempted for ``DF3sum6'' and ``DF3hyp''. A ``perfect'' fit
(2 fit parameters, 2 data points) can be found for P$_-$ and the P-wave
spin-orbit splitting, if both scalar and vector linear potentials are used.
However, all P- and D-wave data [P$_-$, SOS(1P), D$_-$, SOS(1D)] can not be
fitted using the two linear rising potentials and adding a scalar term
$m\omega L(L+1)$ still does not give  a good $\chi^2$.
}
\label{DiracTable}
\end{table}

\section{Conclusions}

\begin{itemize}
\item
With the ``DF3hyp'' lattice, our predictions for the $1^+$ and $2^+$ P-wave state
masses agree very well with the experimental results. We also predict that the
masses of the two lower P-wave states ($0^+$ and $1^+$) should lie only a few MeV
below the $BK$ and $B^\ast K$ thresholds respectively.

\item
Also with the ``DF3hyp'' lattice, the P-wave spin-orbit splitting is small (essentially
zero), but the D-wave spin-orbit splitting is clearly non-zero and positive.
In contrast, another lattice group finds the P-wave spin-orbit splitting to be positive
(about 35~MeV) and the D-wave SOS to be slightly negative (see~\cite{II}), i.e. they
seem to observe the famous inversion \cite{Schnitzer}. However, the recent  European
Twisted Mass Collaboration results find the P-wave SOS to be negative and
the D-wave SOS to be small \cite{TwistedMass}. One clearly needs to go to the continuum
limit before any definite conclusions can be made.

In~\cite{LeeLee}
Woo Lee and Lee suggest that the absence of spin-orbit inversions can be explained by
chiral radiative corrections in the potential model. Small spin-orbit splittings throughout
the meson spectrum could be explained by a relativistic symmetry in the Dirac Hamiltonian
discussed in~\cite{Page}. This would indicate that the scalar potential is (at least
approximately) equal to the vector potential.

\item
The one-body Dirac equation model with one-gluon exchange, vector and scalar linear
potentials and a scalar term $m \omega L(L+1)$ (like a flux tube rotational energy) is
not good enough to describe the entire lattice energy spectrum. Therefore, one should be
very careful in using such simple potentials to describe the interaction between quarks.

\end{itemize}

\section*{Acknowledgements}

I am grateful to my  collaborators, Professors A.M. Green and  C. Michael, and to
the UKQCD Collaboration for providing the lattice configurations.
I wish to thank Professor Philippe de Forcrand for useful comments
and also the Center for Scientific Computing in Espoo,
Finland, for making available the computer resources.
This work was supported in part by the EU Contract No. MRTN-CT-2006-035482,
``FLAVIAnet''.

\appendix

\section{Some checks}

\begin{figure}
\centering
\includegraphics[height=0.51\textwidth, angle=-90]{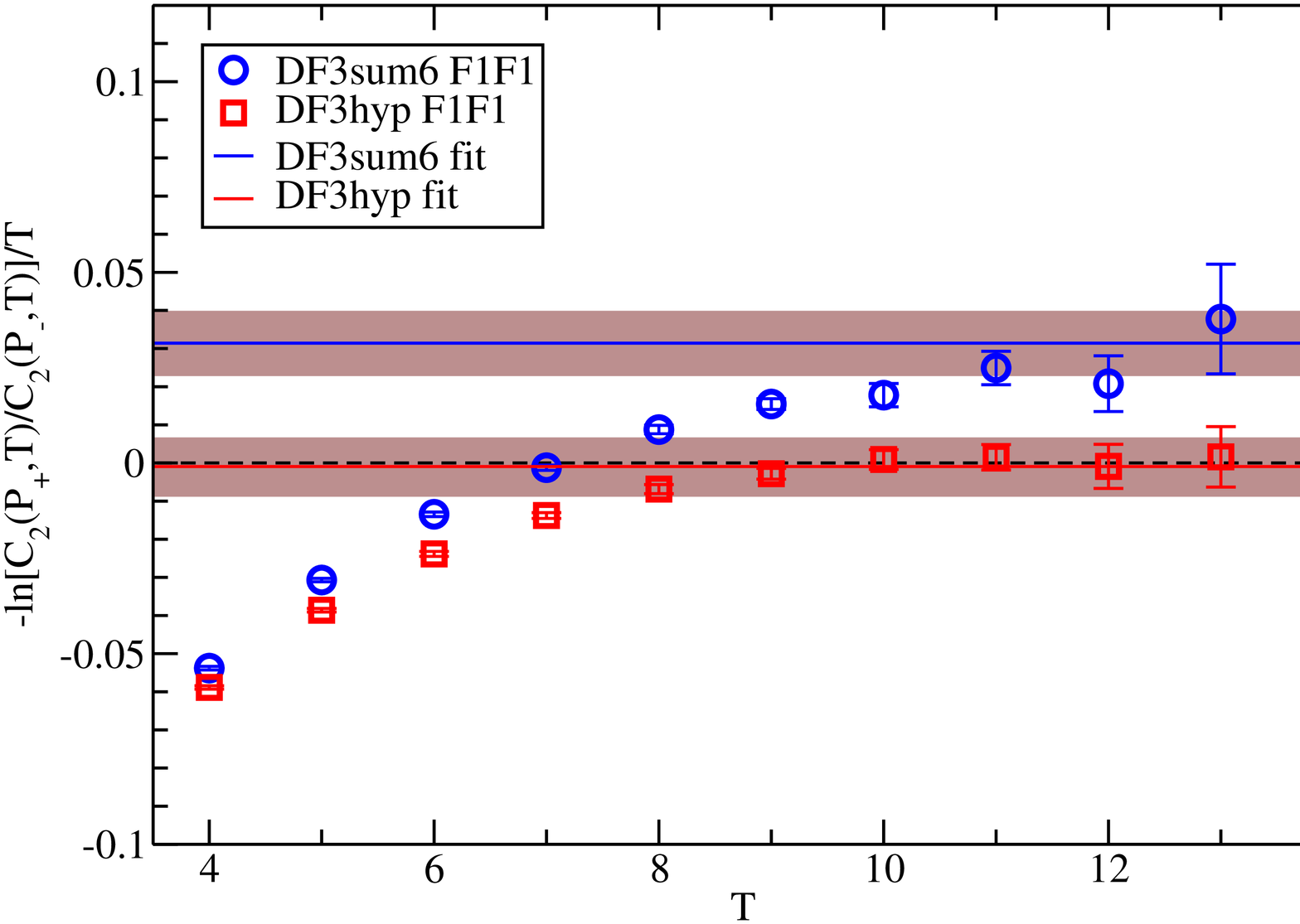}
\includegraphics[height=0.51\textwidth, angle=-90]{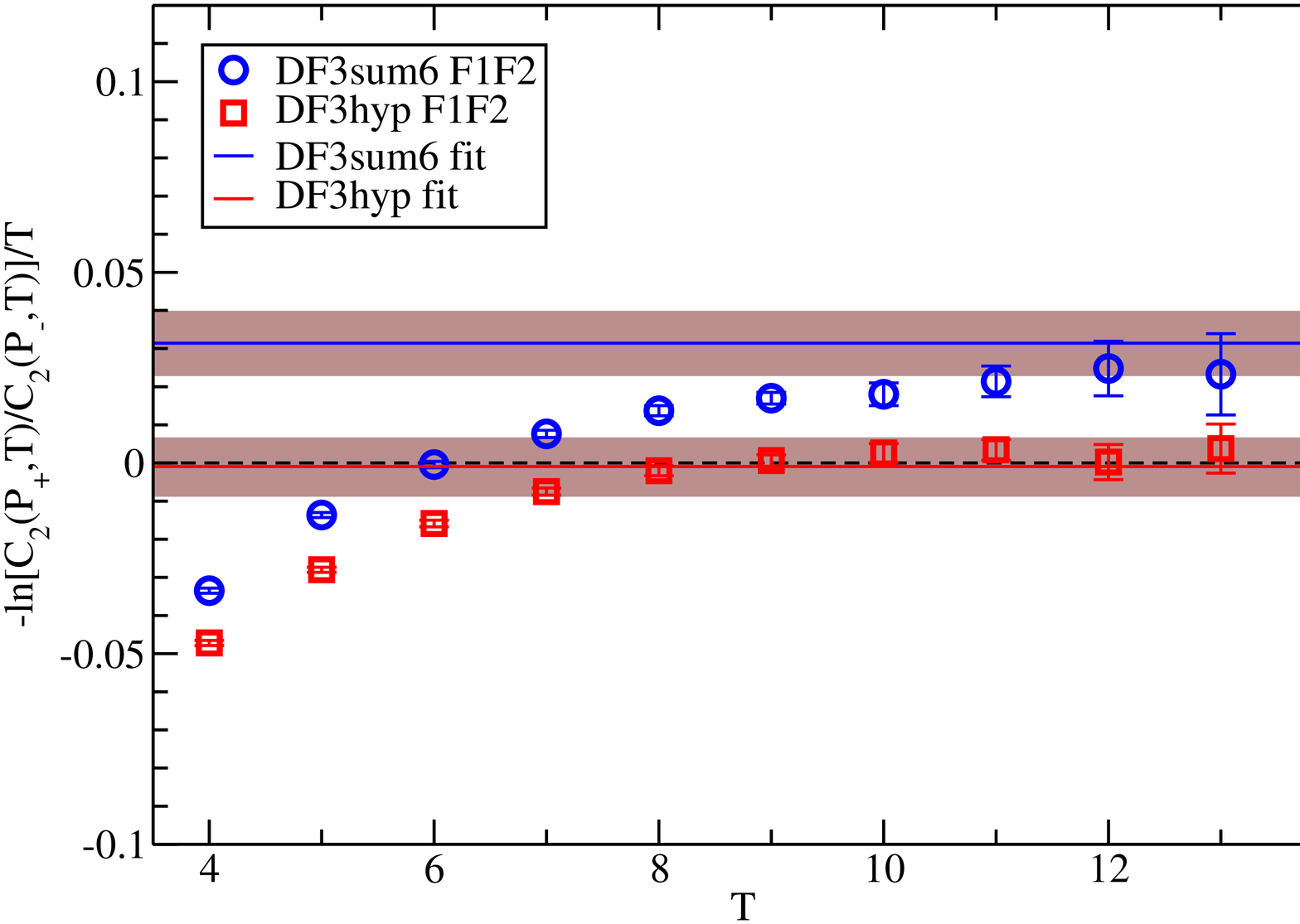}
\includegraphics[height=0.51\textwidth, angle=-90]{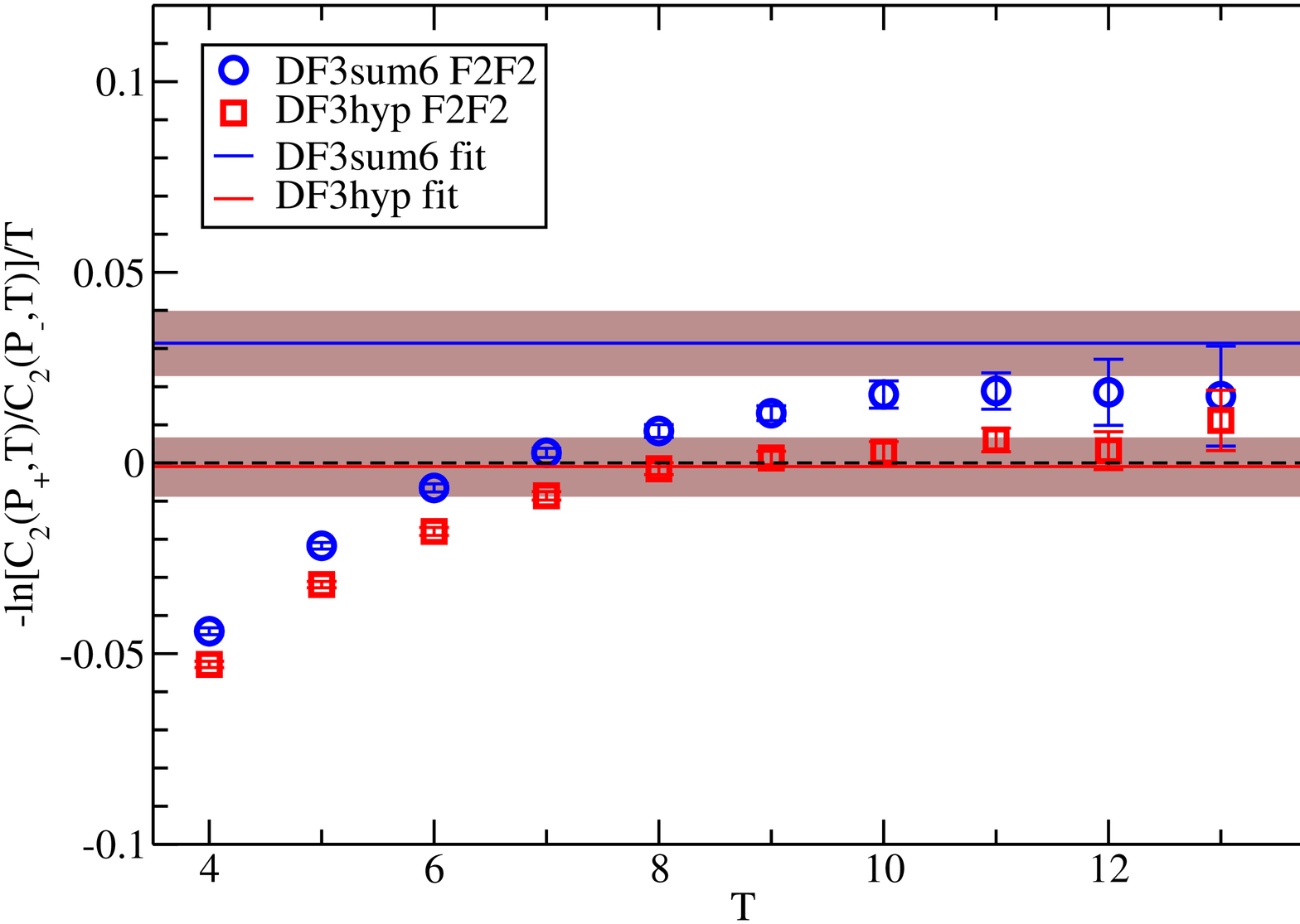}
\caption{(Color online) To study the observed difference in the P-wave
spin-orbit splitting between the ``DF3hyp'' and ``DF3sum6'' configurations
seen in Fig.~\ref{fig:SOSPDF3}, these figures show the logarithm of the ratio
$C_2(\textrm{P}_+,T)/C_2(\textrm{P}_-,T)$ divided by $T$ for ``DF3hyp'' and
``DF3sum6'' for  the three possible fuzzing combinations (from top to bottom:
F1F1, F1F2 and F2F2). This should exhibit a plateux for large enough $T$. The
horizontal lines show the fit results from the full analysis (F1F1, F1F2 and
F2F2 all included). The shaded area shows the estimated errors.
}
\label{fig:SOS_hyp_vs_sum6}
\end{figure}

To check that the different results for different smearings (specifically for
the spin-orbit splittings) are a real effect and not due to some biasing
element in the final analysis, we plot the basic signals in
Figs.~\ref{fig:SOS_hyp_vs_sum6} for the P-wave.
The signals clearly are different for ``DF3sum6'' and ``DF3hyp'',
supporting the results of the complete analysis in Fig.~\ref{fig:SOSPDF3}.

\end{document}